\documentclass[aps,pre,twocolumn,groupedaddress,showpacs,amsmath,amsfonts,10pt,%
floatfix,nofootinbib]{revtex4}
\usepackage{graphicx}
\usepackage{amsmath}
\usepackage{amssymb}

\usepackage{color,colordvi}
\definecolor{Red}{rgb}{0.9,0.0,0.1}

\definecolor{Green}{rgb}{0.1,0.6,0.1}

\begin{document}

%Title of paper
\title{Stiff directed lines in random media}

\author{Horst-Holger Boltz}
\email[]{horst-holger.boltz@udo.edu}

\author{Jan Kierfeld}
\affiliation{Physics Department, TU Dortmund University, 
44221 Dortmund, Germany}

\date{\today}

\begin{abstract}
We investigate the localization of stiff directed lines
 with bending energy by a short-range random potential. 
We apply perturbative arguments, Flory scaling arguments,  a variational 
replica calculation, and functional renormalization to show that a stiff
 directed  line in 1+d  dimensions
undergoes a localization transition 
with increasing disorder  for $d>2/3$. 
We demonstrate that this transition
 is  accessible by numerical transfer matrix calculations
 in 1+1 dimensions and analyze 
the properties of the disorder dominated phase in detail. 
On the basis of the two-replica problem, we propose a relation between 
 the localization of stiff directed lines in 1+d 
dimensions and of  directed lines under tension 
in 1+3d  dimensions, which is strongly supported by identical free 
energy distributions. 
This shows  that  pair interactions in the replicated 
Hamiltonian  determine the nature of directed line
 localization transitions
with consequences for the critical behavior of the 
Kardar-Parisi-Zhang (KPZ) equation. We support the proposed
relation to directed lines via multifractal analysis revealing an analogous
Anderson transition-like scenario and a matching correlation length exponent.
Furthermore, we quantify how the persistence length 
of the stiff directed line is reduced by disorder.
\end{abstract}


%02.50.-r Probability theory, stochastic processes, and statistics 
%05.40.-a Fluctuation phenomena, random processes, noise, and Brownian motion 
%64.70.-p 	Specific phase transitions
%64.60.Ht 	Dynamic critical phenomena
%61.41.+e 	Polymers, elastomers, and plastic

\maketitle

%%%%%%%%%%%%%%%%%%%
\section{Introduction}

Elastic manifolds in random media, especially the 
 problem of a directed
line (DL) or directed polymer in a random potential, are one of the most
important 
model systems in the statistical physics of disordered systems
\cite{HalpinHealy1995}. 
DLs in random media are related to  important 
non-equilibrium statistical physics problems 
such as stochastic growth, in particular 
the Kardar-Parisi-Zhang (KPZ) equation \cite{Kardar1986},
 Burgers turbulence, or the asymmetric 
simple exclusion model (ASEP) \cite{Krug1997}.
 Furthermore, there are many and important 
applications of DLs in random media 
such as  kinetic roughening \cite{Krug1997},
pinning of flux lines in type-II superconductors 
\cite{Blatter1994, Nattermann2000}, 
domain walls in random magnets, 
or wetting fronts \cite{HalpinHealy1995, Yunker2013}. 

Directed lines have a preferred direction and no overhangs 
with respect to this direction. 
The energy of DLs such as flux lines, domain walls, wetting fronts 
is proportional to their length; therefore, the elastic properties
of directed lines are governed by their line tension,
which  favors the straight 
configuration of shortest length. 
Both thermal fluctuations and a short-range random potential (point disorder) 
tend to roughen the DL against the line tension. 
As a result of the competition between thermal fluctuations and disorder,
DLs  in a random media in $D=1+d$ dimensions exhibit a
disorder-driven localization transition \cite{Imbrie1988} for dimensions
$d>2$, i.e., above a critical dimension $d_c=2$. 
These transitions have been 
studied numerically for dimensions up to  $d=4$ 
\cite{Derrida1990,Kim1991,Kim1991b,Monthus2006a,Monthus2006b,Monthus2006c,
Monthus2007,Monthus2007b, Monthus2007c,Monthus2008,Schwartz2012}.
At low temperatures, the DL is in a disorder dominated phase and localizes 
into a path optimizing the random potential energy and the tension 
energy. Within this disorder dominated phase, the DL roughens, there are 
macroscopic energy fluctuations and a finite pair
overlap between replicas 
\cite{Mezard1990,Mukherji1994,Comets2003} (introduced below).
At high temperatures, 
the disorder is an irrelevant perturbation,
and the DL exhibits essentially thermal fluctuations against the 
line tension. 
It has been suggested that the critical temperature
for the localization transition of a DL in a random medium
 and the binding of two DLs by a short-range attractive potential 
coincide \cite{Monthus2006a, Monthus2007}.

DLs in a random medium map onto the dynamic KPZ equation for 
nonlinear stochastic surface growth with the 
 restricted 
free energy of DLs in 1+d dimensions satisfying the KPZ equation 
of a $d$-dimensional dynamic interface. 
The localization transition of DLs with increasing disorder 
corresponds to a roughening transition of the KPZ interface 
with increasing nonlinearity. 
In the context of the KPZ equation, it  is a long-standing 
open question (recently discussed for example in Ref.\ \cite{Pagnani2013})
whether there exists an upper critical dimension,
where the critical behavior at the localization transition is modified.
Therefore, the critical behavior of lines in random media can 
eventually also shed light 
onto the critical properties of the KPZ equation.

\begin{figure}
  \includegraphics[width=0.45\textwidth]{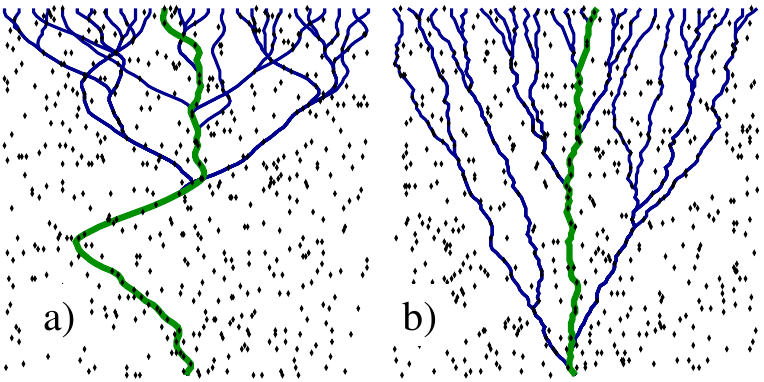}
 \caption{(color online) Paths with lowest energy for given ending states
   (top, globally optimal path thicker) for (a) the stiff  and (b) 
    the tense
   directed line (right). %The minimization of the energy corresponds to 
%$T=0$. 
The dots mark  favorable
   regions of the Gaussian random potential $V$
  (realizations of  the quenched
   disorder are not identical in a and b).} 
 \label{fig:t0}
\end{figure}

In the present paper, 
we study the localization transition of {\em stiff} directed
lines (SDLs). We define SDLs as directed lines with 
 preferred orientation and  no overhangs with respect to this
direction, but with a different elastic energy as compared to DLs: 
SDLs are governed by bending energy, 
which penalizes curvature, rather than line tension, which 
penalizes stretching of the line. This gives rise to configurations 
which are locally curvature-free, i.e., straight but 
straight segments can assume any orientation even if this increases
the total length of the line. 
 We investigate the disorder-induced
localization transition of SDLs for a short-range random potential
 and the  scaling properties of conformations 
 in the disordered phase.
A typical optimal SDL configuration in the presence of 
an additional short-range random potential at zero temperature 
is  shown in  Fig.\ \ref{fig:t0}(a), in comparison to a  typical optimal 
DL configurations in  Fig.\ \ref{fig:t0}(b).

There are a number of applications for SDLs in random media. 
SDLs describe  semiflexible polymers smaller than their
persistence length, such that the assumption of a directed line 
is not violated. Our results apply to 
semiflexible polymers such as DNA or cytoskeletal filaments like F-actin
in a random environment
as it could be realized, for example, by  a porous medium
\cite{Dua2004}.
Moreover, SDLs are closely connected to 
surface growth models for  molecular beam epitaxy (MBE) \cite{Barabasi1995}.
In the presence of surface diffusion, MBE can be described by 
a  Herring-Mullins linear diffusion
equation\cite{DasSarma1994, Racz1994}, which is 
equivalent to the overdamped equation of motion of a SDL.
As a result, SDLs exhibit the same super-roughness as 
MBE interfaces. 
More generally, the zero temperature problem of a
SDL in disorder can be formulated   as a generic  optimization 
problem for paths in an array of randomly distributed 
favorable ``pinning'' sites (as indicated by points in 
Fig.\  \ref{fig:t0}), which minimize their bending energy 
at the same time as  maximizing  the  number of 
visited favorable pinning sites.

There is an important relation between the statistical physics of 
DLs and SDLs, which stems from the return probabilities of {\em  pairs} 
of lines: the contact probability $p_{\rm contact}(L)$ 
of  two thermally fluctuating 
DLs in 1+3d dimensions, i.e., the probability of two DLs  with common 
starting point to meet again after length $L$, 
decays with the same power law $p_{\rm contact}\sim L^{-3d/2}$ as the 
contact probability of SDLs in 1+d dimensions 
\cite{Bundschuh2000,Kierfeld2005}.
We will provide strong numerical evidence  for $d=1$ 
 that this relation between DLs in 1+3d dimensions
and SDLs  in 1+d dimensions not only  holds for purely thermal 
fluctuations but also  in the presence of a  random medium
with a short-range disorder potential.
This relation then allows to 
 address the localization transition of DLs from a different
perspective\cite{Kierfeld2005}.
 Because the proposed relation between DLs and SDLs is based on 
return probabilities of {\em  pairs} of replicas our results 
also  suggest that the 
critical  properties of DLs in a random potential and, thus,
the critical properties of the KPZ equation 
are determined by  the corresponding 
two replica problem.

In particular, the relation between DLs and SDLs implies
that the critical dimension for the localization of SDLs
is $d_c=2/3$ rather than $d_c=2$ for DLs. 
Therefore, SDLs exhibit a  localization transition 
already in $D=1+1$ dimensions, which is easily amenable to numerical 
transfer matrix studies, whereas it requires numerical studies 
 in $D=1+3$ dimensions to investigate the localization transition 
of DLs. 
Therefore, we can verify several  concepts that have been
proposed or found for the localization transition of DLs, e.g.,
 the pair overlap as order parameter or the
multifractality of the localization transition, by simulations 
of SDLs in $D=1+1$ dimensions.

The paper is structured as follows. In the next section, we introduce the
model for stiff directed lines (SDLs) and comment on its relation to directed
lines (DLs). The paper is then divided into two
parts, analytical considerations and numerical findings. First, we apply
scaling analysis to determine the lower critical dimension for the
localization transition as well as estimates for the roughness in the
localized phase. Then we present analytical treatments using variation in
replica space and functional renormalization group and outline difficulties
associated with both techniques. At last, we introduce a numerical transfer
matrix algorithm for a SDL in a random medium and present numerical
results, which validate the existence of a disorder-driven phase transition  
into a localized, roughened phase at low temperatures. We present numerical
results for the roughness, the disorder-induced persistence length, 
Derrida-Flyvbjerg singularities, and the free energy distribution. Furthermore,
we investigate numerically the pair overlap order parameter and multifractal
properties of the transition.    
 
A short account
of some of these results has already appeared as a Rapid Communication
\cite{Boltz2012}.

%%%%%%%%%%%%%%%%
\section{Model of the stiff directed line.}

The configuration of a general directed line, i.e., one without overhangs or
loops and without inextensibility constraint, 
can be parametrized by $(x,{\bf z}(x))$
$(0\leq x\leq L)$ with a $d$-dimensional displacement 
 ${\bf z}(x)$  normal to its preferred orientation. 
In the following, we call the fixed projected length $L$ the length 
of the line. The contour length of the line is given by 
$L_c \approx L(1+\int_0^L dx \partial_x {\bf z}(x))^2/2)$ 
(to leading order in ${\bf z}$); it is not fixed and always larger 
than the length $L$. 
Each configuration of a SDL is associated with an  energy 
\begin{align}
 {\cal H} &= 
  \int_0^L \mathrm{d}x\,\left[\frac{\kappa}{2} (\partial_x^2 {\bf z}(x))^2 
          + V(x, {\bf z}(x))\right], 
\label{eq:ham}
\end{align}
where the first term is the bending energy (to leading order in ${\bf z}$). 
The second term is the disorder energy with a 
Gaussian distributed 
quenched  random potential $V(x,{\bf z}(x))$ 
 with zero mean $\overline{V}=0$ and
short-range correlations
\begin{align}
\overline{V(x,{\bf z})V(x',{\bf z}')} = 
  g^2\delta^d({\bf z}-{\bf z}')\delta(x-x')  \label{eq:corr}
\end{align}
$\overline{X}$ denotes the quenched disorder average over
  realizations of $V$, whereas ${\langle}X{\rangle}$ denotes  thermal
  averaging.

The SDL model \eqref{eq:ham} is often used as a
weak-bending  approximation
to the so-called worm-like chain or Kratky-Porod
model \cite{Harris1966,Kratky1949} 
\begin{align}
{\cal H}_{\text{WLC}} &=
\int_0^L\mathrm{d}s\, \left[\frac{\kappa}{2} (\partial_s^2 {\bf r}(s))^2 
          + V({\bf r}(s))\right]
\end{align}
which is the basic  model for inextensible semiflexible polymers, 
such as 
DNA or cytoskeletal filaments like F-actin. The chain is
parametrized in arc length, leading to a local inextensibility constraint
$\lvert \partial_s {\bf r}(s)\rvert=1$, which is hard to account for, both
numerically and analytically \cite{Kleinert2006,Dua2004}. 
For thermally  fluctuating semiflexible
polymers,
the approximation \eqref{eq:ham} only applies to a weakly bent  
semiflexible polymer on length scales
{\em below} the so-called persistence length,
 which is \cite{Kleinert2006,Gutjahr2006}
\begin{equation}
L_p =\frac{2}{D-1}\frac{\kappa}{T} = \frac{2}{d}\frac{\kappa}{T}. 
\label{eq:lp}
\end{equation}
in $D=1+d$ dimensions. 
We use here and throughout the following energy
units with $k_B\equiv 1$.
Above the persistence length, a semiflexible polymer 
loses orientation correlations and starts to develop 
overhangs. 

Also in a quenched random potential the SDL model
describes semiflexible  polymers 
in heterogeneous media, only  as long as 
tangent fluctuations are small such that overhangs can be
neglected, which  is the case below a {\em disorder-induced} persistence 
length, which we will derive below.

We consider the SDL model also in the
thermodynamic limit beyond this persistence length, 
because  we find evidence for  a relation to 
 the important problem of DLs in a random medium 
in {\em lower} dimensions. This relation 
is based on replica pair interactions and 
shows that pair interactions also determine the nature of the
DL localization transition. Moreover, this relation can make
the DL transition in high dimensions  
computationally accessible. 
  We will now outline the idea behind 
this relation.

%%%%%%%%%%%%%%%%%%%%%%%
\section{Relation to directed lines}

\begin{figure}[t]
 \begin{center}
  \includegraphics[width=0.45\textwidth]{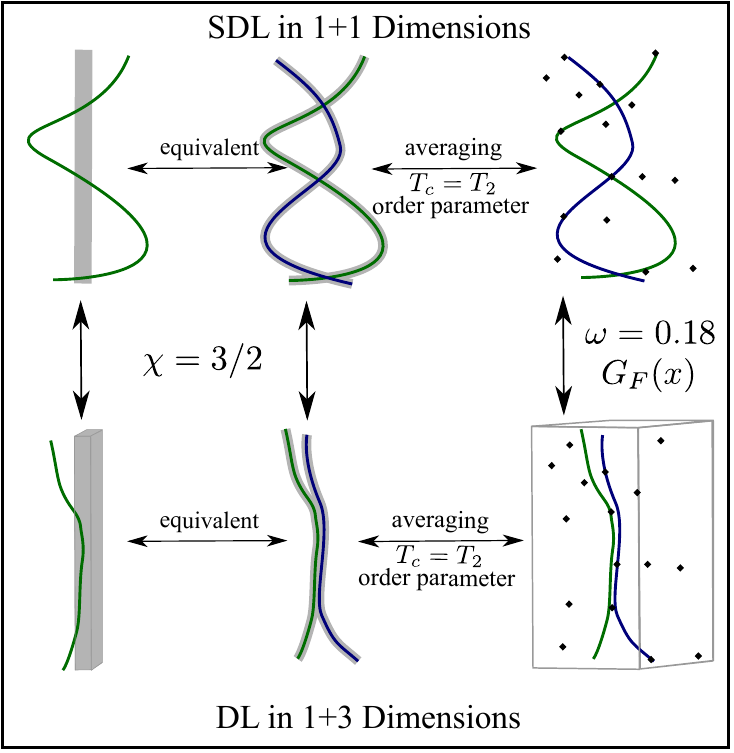}
 \caption{Schematic summary of our findings for the relation
between DLs and SDLs. Our results for the energy fluctuation exponent
$\omega$ (introduced below in eq.\ (\ref{eq:scalrel})) and for the free energy
distribution $G_F(x)$ are presented in sections \ref{sec:rough} and
\ref{sec:distrib}. The shaded area marks the range of the short-ranged (binding)
potential. The effective binding interaction becomes apparent in the replica
formalism (section \ref{sec:replica}); the importance of two-replica
interactions for SDLs is one essential result of this work, see sections
\ref{TcT2} and \ref{sec:overlap}.}\label{fig:result2}
 \end{center}
\end{figure}

The  difference between  SDLs and   DLs
  \cite{HalpinHealy1995} is the {\em second} derivative in the
first bending energy term in eq.\
(\ref{eq:ham}) for SDLs, which differs from  the  tension
or stretching energy
$\sim \int dx \frac{\tau}{2} (\partial_x {\bf z}(x))^2$ of DLs
with line tension $\tau$. 
This results in  different types of energetically favorable configurations
 (see Fig.\ \ref{fig:t0}): large perpendicular 
displacements ${\bf z}$ of SDLs as shown in  Fig.\ \ref{fig:t0}(a) are not 
unfavorable as long as their ``direction'' does not change, i.e., 
as long as their are locally straight and, therefore, 
do not cost bending energy. Such configurations increase, however, 
the length of the line and are suppressed for DLs by the 
tension or stretching energy. 

The statistics of displacements ${\bf z}$ is characterized by the 
 {\em roughness exponent} $\zeta$, which is  defined by
$\overline{\langle z^2(L)\rangle}\sim L^{2\zeta}$.
The  thermal roughness is   $\zeta_{th,\tau}=1/2$ for  DLs and
$\zeta_{th,\kappa}=3/2$ for SDLs: equating the thermal energy $T$ 
with the respective elastic energies gives $T\sim \tau z^2/L$ 
for DLs and $T\sim \kappa z^2/L^3$ for SDLs.
Here and in the following we
use subscripts $\tau$ (tension) and $\kappa$ (bending stiffness) 
to distinguish between the two systems.

Although typical configurations are quite different,
a SDL subject to a short-ranged 
(around $z=0$) attractive potential
$V({\bf z})$ can be mapped onto a DL in high dimensions
$d'=3d$ \cite{Bundschuh2000,Kierfeld2005}. 
This equivalence is based on the probability that a
free line of length $L$ starting at
(${\bf z}(0)=0$) ``returns'' to the origin, i.e., ends at ${\bf z}(L)=0$.
This return probability is characterized by a {\em return  exponent} $\chi$: 
$\text{Prob}({\bf z}(L)=0)\sim L^{-\chi}$.
The same return exponent characterizes  the contact probability 
$p_{\rm contact}(L)\sim L^{-\chi}$, i.e., the probability 
 of two lines  with common 
starting point to meet again after length $L$, as follows from 
considering the relative coordinate. 
For DLs, which are  essentially random walks in $d$ transverse
dimensions, the return exponent is $\chi_{\tau}=d/2$ \cite{Fisher1984}, 
whereas it is $\chi_{\kappa}=3d/2$ for a SDL (after 
integrating over all orientations of the end) \cite{Gompper1989}; 
they are related to the roughness exponents by $\chi= \zeta d$
\cite{Kierfeld2005}.
The return exponent  $\chi$ 
governs the critical properties of the 
binding transition to a short-range attractive potential 
or, equivalently, of the binding transition 
of two lines interacting by such a potential
\cite{Bundschuh2000, Kierfeld2005}.
This follows, for example, directly from a necklace model
treatment \cite{Fisher1984}.
The relation 
\begin{equation}
  \chi_{\tau}(3d)=\chi_{\kappa}(d)
  \label{eq:mapping}
\end{equation}
implies that the binding transition 
of two DLs in $3d$ dimensions maps onto 
to the binding transition of two SDLs in $d$ 
dimensions.

In the replica formulation of line localization problems such as 
 (\ref{eq:ham}),
the random potential gives rise to a short-range 
attractive pair interaction (see below). 
Therefore, pairwise interactions of DLs
play a prominent role also for the physics of a single 
DL in a random potential. 
Furthermore, the critical temperature
$T_{c,\tau}$ for a DL in a random potential is
believed to be identical to the critical
temperature $T_{2,\tau}$ for a system with two replicas
\cite{Monthus2006a, Monthus2007}. In section \ref{TcT2},
we will show numerically
that also for SDLs in a random medium $T_{c,\kappa}=T_{2,\kappa}$ holds.
The important role of pairwise interactions suggests that 
 not only the binding transition 
of two DLs in $3d$ dimensions maps onto 
to the binding transition of two SDLs in $d$ 
dimensions but that the same dimensional relation 
holds for  the localization transitions of DLs and SDLs in 
a short-range random potential.

One aim of this
work is to support this conjecture by providing strong numerical 
evidence 
that the $d\rightarrow 3d$ analogy between DLs and SDLs in 
a random potential holds for the {\em entire}
free energy distributions for DLs in 1+3 and SDLs in 1+1 dimensions. 
Because this analogy is rooted in 
the binding transition of replica pairs, we can conclude that
 critical properties of the localization transition are 
determined by  pair interactions in the replicated 
Hamiltonian, which  has  been previously 
suggested in Refs.\ \cite{Bundschuh1996,Mukherji1996}. 

Moreover, it has been proposed  that 
pair interactions can be used to formulate an 
 order parameter of the disorder-driven localization 
transition of DLs in terms of the overlap 
$q\equiv \lim_{L\rightarrow\infty}\overline{\frac{1}{L}\int_0^L\mathrm{d}x
\delta(z_1(x)-z_2(x))}$ \cite{Mezard1990,Mukherji1994,Comets2003}, i.e.,
 the average number of sites per length, that two lines in the same
realization of the disorder have in common.
Localization by disorder gives rise to a 
 finite value of the pair overlap $q$. This coincides with the binding energy
per length that characterizes binding of 
two polymers by a pair potential.
We will also show that the pair overlap indeed provides 
a suitable order parameter for the localization transition 
of SDLs in 1+1 dimensions. A schematic summary of the relation of
DLs and SDLs together (together with some relevant numeric results) is shown
in Fig.\ \ref{fig:result2}.

%%%%%%%%%%%%%%%%%%%%%%%%%%%%%%%
\section{Scaling analysis}

\begin{figure}[t]
\begin{center}
 \includegraphics[width=0.35\textwidth]{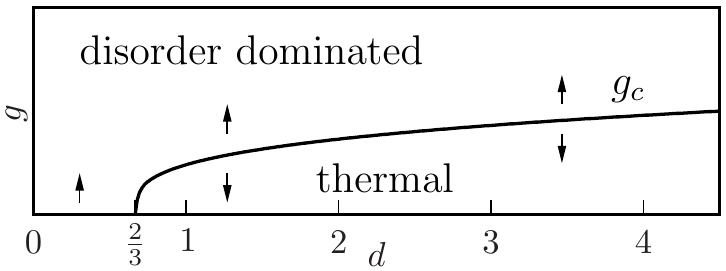}
 \caption{Sketch of the phase diagram as implied by Flory type arguments.
The arrows indicate the flow of the disorder ``strength'' $g$ under
renormalization.}\label{fig:renorm}
\end{center}
\end{figure}

%%%%%%%%%%%%%%%%%%%%%
\subsection{Lower critical dimension}

We start with a scaling analysis by a 
 Flory-type argument. For displacements $\sim z$
the bending energy  in eq.\
 (\ref{eq:ham}) scales as $E_b \sim z^2 L^{-3}$, which also leads to 
$\langle z^2 \rangle \sim  L^3/L_p$ and the thermal roughness exponent
$\zeta_{th,\kappa}=3/2$.
The disorder energy in eq.\  (\ref{eq:ham})
 scales as $E_d \sim \sqrt{L z^{-d}}$, 
as can be seen from eq. \eqref{eq:corr}. 
Using the  unperturbed  thermal roughness in the disorder energy 
we get  $E_d \sim L^{(2-3d)/4}$, 
from which we conclude that the disorder
is relevant below a {\em lower critical dimension} $d_{c,\kappa}$ with 
\begin{equation}
   d_{c,\kappa}=2/3.
\end{equation} 
Above this critical dimension for $d>2/3$ and, thus, 
in all physically accessible integer 
dimensions, the SDL should exhibit a transition 
from  a thermal phase for low $g$  to a disorder dominated phase above a 
critical value $g_c$ of the disorder (see Fig.\ \ref{fig:renorm}). Of course
the same distinction could be made in terms of the temperature with a 
high temperature phase for $T>T_c$ 
and a disorder dominated low temperature phase
for $T<T_c$. In the disorder dominated phase, the SDL becomes localized or
``pinned'' into a configuration favored by the random potential and assumes a
roughened configuration, see Fig.\ \ref{fig:t0}. 
We remember that the lower critical dimension for 
DLs is $d_{c,\tau}=2$, which is in accordance with the relation 
between  DLs in 1+3d dimensions
and SDLs  in 1+d dimensions proposed in the previous section.
We also point out  that for both types of lines the return 
exponent $\chi$ assumes the universal value 
$\chi = 1$ at the critical dimension 
because of  $\chi_{\tau} = d_{c,\tau}/2=1$ and 
$\chi_{\kappa}=3d_{c,\kappa}/2=1$ \cite{Bundschuh2000,Kierfeld2005}. 

With  $d_{c,\kappa}=2/3<1$
 the localization transition of SDLs  can be 
studied numerically in 1+1 (or higher) integer dimensions,
whereas for DLs with  $d_{c,\tau}=2$,
 the localization transition is  only accessible in 
simulations in  1+3 (or higher) integer dimensions.
Our numerical study of the line localization transition 
presented below focuses on SDLs in 1+1 dimensions, which are 
computationally better  accessible using transfer matrix techniques
as compared to a 1+3 dimensional system of DLs. 
This allows us to verify important concepts for the localization 
transition such as the overlap order parameter in section 
\ref{sec:overlap}, which have not been accessible for 
DLs in 1+3 dimensions up to now.

%%%%%%%%%%%%%%
\subsection{Roughness}

Balancing the Flory estimates, $E_b \sim E_d$,  gives a roughness 
estimate  $z\sim L^{\zeta_{Fl}}$. When disorder is relevant, this leads to 
\begin{equation}
 \zeta_{Fl,\kappa}=\frac{7}{4+d}~~\mbox{for}~~d<d_{c,\kappa}=\frac{2}{3}.
\end{equation}
This result is only applicable below the critical dimension $d<d_{c,\kappa}$,
where $\zeta_{Fl,\kappa}>\zeta_{th,\kappa}$.
Above the critical dimension, the Flory-result would give  
 $\zeta_{Fl,\kappa}<\zeta_{th,\kappa}$, which  contradicts 
our expectation that the SDL roughens as it 
 adjusts to the random potential. 
Furthermore, the exponent $\omega$ related to the sample to sample
free energy
fluctuations $\Delta F^2 \equiv \overline{(F-\overline{F})^2}$
via $\Delta F \sim L^{\omega}$
would be negative because of the  general
scaling relation
\begin{align}
\omega=2\zeta_{\kappa}-3.
 \label{eq:scalrel}
\end{align}
This scaling relation follows from the scaling of the 
bending energy $E_b \sim z^2 L^{-3} \sim L^{2\zeta_{\kappa}-3}$ and 
the assumption $\Delta F \sim E_b$ that bending and free energy
 have the same scale dependence.
Note that we do not subscript $\omega$ in eq.\ (\ref{eq:scalrel})
as we believe that 
$\omega_{\tau}=\omega_{\kappa}$, see section \ref{sec:numericalresults}.
An exponent $\omega<0$  contradicts  the existence of  large disorder-induced
free energy fluctuations in the low-temperature phase
\cite{Fisher1991,Monthus2006b}, for which there is also strong numerical
evidence\cite{Kim1991,Kim1991b,Monthus2006a}.
We conclude that this kind of argument is not applicable above the critical
dimension. The same problems  occur in Flory arguments  for the DL
for $d>d_{c,\tau}=2$, where one finds  $\zeta_{Fl,\tau}=3/(4+d)$ and 
$\omega = 2\zeta_{\tau}-1$.
The Flory approach underestimates  the roughness exponent 
for all dimensions both for DLs and SDLs, i.e., $\zeta > \zeta_{Fl}$ as it
features only one length scale on which the line can adjust 
to the disorder. Therefore the Flory results $\zeta_{Fl}$ should provide 
a lower bound for the roughness exponent, i.e., 
$\zeta> \zeta_{Fl}$. Moreover, the thermal roughness  
$\zeta_{th,\tau}=1/2$ for  DLs and
$\zeta_{th,\kappa}=3/2$ for SDLs should provide another lower bound for the 
disorder-induced roughness, i.e., $\zeta > \zeta_{th}$ because 
of the existence of disorder-induced
 free energy fluctuations in the low-temperature phase, i.e., 
$\omega>0$.

Furthermore,  we can give an upper bound for the disorder-induced 
roughness by an argument which relates the line  to the zero-dimensional 
problem of a single particle in a harmonic and a  
 random potential \cite{Engel1993}.
In this argument we  divide the line in its middle into 
two straight and  rigid segments separated by  the mid-point 
$(L/2,z)$. The single (i.e.\ zero-dimensional)
 coordinate $z$ describes the restricted 
shape fluctuations of the two-segment line. 
The bending  energy  of these shape fluctuations  is 
${\cal H}_b({\bf z}) = \frac{1}{2}(\kappa/L^3) z^2$, and the disorder energy 
of the two straight segments is 
${\cal H}_d({\bf z}) = V({\bf z})$  with $\overline{V({\bf z})}=0$ and 
$\overline{V({\bf z})V({\bf z}')} = (2g^2L)\delta^d({\bf z}-{\bf z}')$. 
In   Ref.\ \cite{Engel1993}, a roughness
$\overline{\langle z^2 \rangle} \sim   (g^2L)^{1/2}/(\kappa/L^3) \sim L^{7/2}$
(plus logarithmic corrections) has been obtained for this
zero-dimensional problem with the single degree of freedom $z$
by an Imry-Ma argument and a more detailed replica calculation,
which implies a roughness exponent 
\begin{equation}
 \zeta_{0,\kappa} = \frac{7}{4}
\label{eq:zeta0}
\end{equation}
for  the stiff two-segment line. 
For the directed line, an analogous 
argument gives $\zeta_{0,\tau} = 3/4$.
In the framework of a functional renormalization group (FRG)
calculation both roughness exponents can be written as 
$\zeta_{0} = \epsilon/4$ in a dimensional expansion with the appropriate
$\epsilon=4-d$ for DLs and $\epsilon=8-d$ for SDLs
 (see  section \ref{sec:FRG} below). 
Using the scaling relations (\ref{eq:scalrel}), 
this results in $\omega_0=\frac{1}{2}$ for two-segment lines
(both for  SDLs and  DLs), which should be
considered an upper bound to the energy fluctuation exponent,
 because the adaptation
of the line to the potential must not lead to larger fluctuations as 
compared to
the trivial case of summing up random numbers \cite{LeDoussal2004}.
Therefore, also $\zeta_0$ should provide an upper bound 
for the roughness exponent,  i.e., $\zeta < \zeta_{0}$.
The upper bound $\zeta<\epsilon/4$  is also found in the  
FRG calculation \cite{LeDoussal2004}.

All in all, we have obtained bounds
\begin{equation}
   \max(\zeta_{th},\zeta_{Fl}) < \zeta < \zeta_0,
\label{eq:bounds}
\end{equation}
which apply both for SDLs and DLs. For SDLs, this gives a 
relatively small window of possible roughness exponents,
\begin{equation}
   \max\left(\frac{3}{2},\frac{7}{4+d}\right) < \zeta_\kappa < \frac{7}{4},
\label{eq:bounds_SDL}
\end{equation}
which, for example,  limits $\zeta_\kappa$ for SDLs in  1+1 dimensions to 
$1.5 < \zeta_\kappa < 1.75$.

%%%%%%%%%%%%%%%%%%%%%%%%%%%%%%%
\section{Variation in replica space} \label{sec:replica}

To go beyond  scaling arguments we use the replica
technique \cite{Dotsenko2001} following the treatment of directed
manifolds by Mezard and Parisi  \cite{Mezard1991}. For the sake of
convenience we restrict ourselves to $d=1$ throughout this section. We will
comment on how to adapt this to higher dimensions later on.
The quenched average of the free energy
 $\overline{F}=-\beta^{-1} \overline{\ln Z}$
is treated in the  representation 
\begin{equation}
 \ln Z = \lim_{n\rightarrow 0} n^{-1} (Z^n-1)
\end{equation}
 calculating averages $\overline{Z^n}$ of a $n$-times replicated 
system in the limit $n\to 0$. 
For the calculation we introduce an additional parabolic 
potential or  ``mass''
term  $\int dx \mu {\bf z}^2(x)$ in eq.\ (\ref{eq:ham}) as an infrared
regularization and a finite correlation length $\lambda$ in the disorder
correlator,
\begin{equation}
\overline{V(x,z)V(x',z')} = g^2f_\lambda((z-z')^2)\delta(x-x'),
\end{equation}
where we use 
$f_\lambda(x)=\sqrt{2\pi\lambda^2}^{-1}\exp{(-x/(2\lambda^2))}$ to
retain the original $\delta$-correlator (compare eq.\ \ref{eq:corr}) for
$\lambda\approx 0$. 
We write the  replicated and averaged partition function as
$\overline{Z^n}=\prod_\alpha(\int\mathcal{D}z_\alpha)\exp{(-\beta
  {\cal H}_{\text{rep}})}$ with the following  replica Hamiltonian in Fourier
space 
\begin{align}
  {\cal H}_{\text{rep}}&=\frac{1}{2L} \sum_{\alpha=1}^n
     \sum_k (\kappa k^4 + \mu) z_\alpha^2
  \nonumber\\
 &  - \frac{\beta g^2}{2} \sum_{\alpha,\beta=1}^n \int_0^L
  \text{d}x\,f((z_\alpha-z_\beta)^2). \label{eq:H_rep}
\end{align}
As mentioned before, ${\cal H}_{\text{rep}}$ is  related to a pair binding
problem: in the  limit $\lambda \approx 0$ the second
term becomes $- \frac{\beta g^2}{2} \sum_{\alpha,\beta} \int_0^L
\text{d}x\delta(z_\alpha-z_\beta)$, which is an  attractive short-range
interaction of two replicated lines.

We use variation in replica space with the quadratic, i.e.,  Gaussian 
trial Hamiltonian
\begin{equation}
   {\cal H}_V= (2L)^{-1} \sum_k \sum_{\alpha,\beta} z_\alpha
{G}^{-1}_{\alpha\beta} z_\beta
\end{equation}
 with ${G}^{-1}_{\alpha\beta}=(\kappa k^4 + \mu)\delta_{\alpha\beta}
+ {\sigma}_{\alpha\beta}$ with the self-energy  matrix
$\boldsymbol{\sigma}$ providing  variational parameters. 
Extremizing  the lower free energy bound 
 $F\geq F_V+\langle {\cal H}_{\text{eff}}-{\cal H}_V\rangle_V$
with respect to the self-energy matrix $\boldsymbol{\sigma}$ gives
(in the continuum limit $L^{-1} \approx 0$) a self-consistent 
equation for the self-energy matrix \cite{Mezard1991}
\begin{align}
  {\sigma}_{\alpha\beta}&=\begin{cases} 
      -\sum_{\alpha'\neq\alpha} {\sigma}_{\alpha\alpha'} & \alpha=\beta
\\
    -2\beta g^2 \tilde{f}'\left(({\int}\frac{\text{d}k}{2\pi\beta}
({G}_{\alpha\alpha}{+}{G}_{\beta\beta}{-}2{G}_{\alpha\beta})\right)
    & \alpha\neq\beta
\end{cases}
\label{eq:sigma}
\end{align}
with 
\begin{equation}
\tilde{f}_\lambda(x)\equiv 
 \int dy \sqrt{2\pi}^{-1} e^{-y^2/2} f_\lambda(y^2x) = 
\sqrt{2\pi}^{-1} \sqrt{\lambda^2+x}^{-1}.
\end{equation} 

Following
 Mezard and Parisi
 we choose a one-step hierarchical replica symmetry breaking Ansatz for
$\boldsymbol{\sigma}$. Replica symmetry breaking is relevant here, 
 because there is no nontrivial replica symmetric solution in the limit 
$\mu \approx 0$ apart from 
$\sigma_{\alpha\beta}=0$.
Furthermore, there is no
continuous replica symmetry breaking since  $\zeta_{Fl}<\zeta_{th}$
for $d>d_{c,\kappa}$ \cite{Mezard1991}. 
Thus, $\boldsymbol{\sigma}$  can be parametrized by a
diagonal element $\tilde{\boldsymbol{\sigma}}$ 
and a function $\sigma(u)$ ($0\leq u
\leq 1)$) giving the non-diagonal elements in the limit $n\rightarrow
0$. For one-step replica symmetry breaking the latter takes the
form $\sigma(u)=\sigma_0+\Theta(u-u_c)(\sigma_1-\sigma_0)$.
Using the algebra developed by Parisi \cite{Parisi1980} and  performing the
limit of an unbounded system $\mu\rightarrow 0$ we find
\begin{align}
\sigma_0 &= 0\\ 
\sigma_1&= -2\beta g^2\tilde{f}'(\pi S_1) 
\end{align}
 with $S_1=S_1(\Sigma) 
\equiv \frac{1}{\sqrt{2}\beta}\kappa^{-1/4}\Sigma^{-3/4}$ and
$\Sigma\equiv u_c \sigma_1$. 
In higher dimensions $d>1$ the stationary equation (\ref{eq:sigma}) 
as well as the result for one-step replica symmetry breaking would be of the
same form with different numerical constants, given the 
functions  $f_\lambda$ and $\tilde{f}_\lambda$ are
adapted accordingly, see \cite{Parisi1980}.
In $d=1$, the variation
of the free energy estimate yields two self-consistent equations
\begin{equation}
 u_c\propto S_1(\Sigma)(\lambda^2+S_1(\Sigma))^{-1}~~~\mbox{and}~~~
 \Sigma \propto  (\beta^{20} g^{16} \kappa^3 u_c^{20})^{-1}
\end{equation}
for $u_c$ and $\Sigma$, 
where we omitted numerical constants. We are not able to give a
closed solution, because
\begin{align}
 \Sigma^{1/20}&\sim \frac{\lambda^2+\Sigma^{-3/4}}{\Sigma^{-3/4}}\\
	      &\sim \lambda^2\Sigma^{15/20}+\text{const}
\end{align}
Nonetheless, using the condition $u_c<1$, we can show 
that there is no solution
unless the potential strength $g$ and correlation length $\lambda$ are
above finite values.
%\begin{align}
% g &> K \beta^{-7/6}\kappa^{-1/6}\lambda^{1/6} \\ 
% 2\lambda^2 &> \tau > 0
%\end{align}
%with a numerical constant $K$.  
Although variation in replica space is known \cite{Mezard1991} to fail in
reproducing the exact solution \cite{Kardar1987} for the problem of the DL in
disorder in finite dimensions, we interpret this as an indication for the
existence of a critical disorder strength or a critical temperature.
The replica approach leads, however, to the thermal roughness exponent 
also in the low-temperature phase.

% %%%%%%%%%%%%%%%%%%%%%%%%%%%%%%%%%%%%%%%
\section{Functional  renormalization group}
\label{sec:FRG}

There has been some success studying elastic manifold problems
in disordered potentials using 
functional  renormalization group (FRG) analysis
\cite{Fisher1986,Balents1993,Chauve2001,LeDoussal2002,LeDoussal2004,
LeDoussal2005}. 
This method
 can be adapted to generalized elastic energies
\begin{equation}
 {\cal H}_m \sim
  \int\!\text{d}^D\!x\,(\nabla^m_{\bf x} {\bf z})^2
\label{eq:HamFRG}
\end{equation}
 for $D$-dimensional 
elastic manifolds with $d$ transverse dimensions 
(in the FRG literature, the number of transverse dimensions is 
frequently denoted by $N$). The case $m=1$  corresponds
 to elastic manifolds 
as they have been already extensively studied
\cite{Fisher1986,Balents1993,Chauve2001,LeDoussal2002,LeDoussal2004,
LeDoussal2005}, whereas
$m=2$ corresponds to  manifolds dominated by bending energy. 
Lines are manifolds with  $D=1$, i.e., 
DLs correspond to $m=1$ and $D=1$ and  SDLs with a bending energy 
to $m=2$ and $D=1$. 
In the FRG approach 
we take the Gaussian distributed random potential 
to have zero mean and a correlator of the 
general form 
\begin{align}
 \overline{V({\bf x},{\bf z})V({\bf x}',{\bf z}')}&=
  R({\bf z}-{\bf z}')\delta^{(D)}({\bf x}-{\bf x}'),
\end{align}
where the whole function $R({\bf z})$  is renormalized under a change 
of scale.

In  renormalization, we integrate out short wavelength fluctuations 
in a shell $\Lambda/b < |{\bf k}| < \Lambda$ in 
momentum space   and  perform a subsequent   infinitesimal
scale-change (SC) by a factor $b=e^{\mathrm{d}l}$, 
\begin{align}
 x&\rightarrow bx\\
 {\bf z}&\rightarrow b^{\zeta}{\bf z},
\end{align}
in order to restore the high momentum cutoff $\Lambda$. 
This leads to the following FRG flow equations
\begin{align}
 \left.\frac{\mathrm{d}T}{\mathrm{d}l}\right|_{\text{SC}} &=
2(\zeta_{th}-\zeta)T
\label{eq:FRGT}\\
 \left.\frac{\partial R}{\partial l}\right|_{\text{SC}}
    &= (\epsilon - 4\zeta)R+\zeta (z\cdot\vec{\nabla}_{z})R +{\cal O}(R^2) + 
    {\cal O}(R^3)
\label{eq:FRGR}
\end{align}
with 
\begin{equation}
   \epsilon=4m-D 
   \label{eq:epsilon}
\end{equation}
 and $\zeta_{th}=(2m-D)/2$\footnote{For $D\geq 2k$ the
manifold does not have a macroscopic roughness and $\zeta_{th}$ can no longer
be interpreted as thermal roughness exponent.}. 
The flow equation (\ref{eq:FRGT}) for the
temperature is believed to be exact due to a Galilean invariance of the
Hamiltonian. For a disorder dominated phase with 
$\zeta>\zeta_{th}$ corresponding to $\omega>0$, the system is 
characterized by  a $T=0$ RG fixed point, at which we want to 
determine the roughness exponent $\zeta$.

The terms ${\cal O}(R^2)$ and ${\cal O}(R^3)$ in the RG flow equation 
(\ref{eq:FRGR}) of the disorder correlation function  $R({\bf z})$ 
 are additional one-loop \cite{Balents1993} and two-loop 
\cite{Chauve2001,LeDoussal2002,LeDoussal2004,LeDoussal2005} contributions, 
respectively. 
In Ref.\ \cite{LeDoussal2004}, a generalized elasticity 
with a general parameter $m$ 
has already been discussed up to two-loop level for $d=1$. 
The one-loop contributions are {\em independent} of $m$ and  assume 
 exactly the same form for $m=2$ as for standard elastic manifolds 
with $m=1$ and as they have been calculated in   Refs.\ 
\cite{Fisher1986,LeDoussal2002,LeDoussal2004} for $d=1$ 
and Refs.\ \cite{Balents1993,LeDoussal2005} for general $d$. 
For $d=1$, it has been shown that the 
 two-loop contributions, however,  acquire  a $m$-dependent 
numerical prefactor \cite{LeDoussal2002,LeDoussal2004}. 
For general $d$ and $m$, the two-loop contribution has not been calculated. 
The exponent $\zeta$ is determined from the 
 FRG equation (\ref{eq:FRGR}) for $R({\bf z})$
by requiring a fixed point solution for short-range disorder 
 to be positive and vanish exponentially for large $z$. 
Therefore, in one-loop order 
results for the roughness exponent $\zeta$ 
depend on $m$ only through the dimensional  expansion 
parameter $\varepsilon=4m-D$.

For $d=1$, we can adapt
the final results achieved in Ref.\ \cite{Fisher1986} in one-loop, which have
been extended in Refs.\ \cite{LeDoussal2002,LeDoussal2004} to two loops,
\begin{equation}
   \zeta_{\rm FRG} = 0.20830\varepsilon + 0.00686\,X_m\,\varepsilon^2
\label{eq:zetaFRG}
\end{equation} 
with the  $m$-dependent 
numerical prefactor $X_m$ ($X_1=1$, $X_2=-1/6$ to leading order in
$\varepsilon$). 
For a SDL 
($D=1$, $m=2$, $\epsilon=7$)  with  $d=1$ transverse dimensions, we obtain
a roughness exponent $\zeta_{{\rm FRG},\kappa}
\approx 1.4571$ in one-loop,
 which is close to the Flory estimate
$\zeta_{Fl,\kappa}=\frac{7}{5}$ but also violates the 
lower bound set by the thermal roughness,
  $\zeta_{{\rm FRG},\kappa}<\zeta_{th,\kappa}=3/2$, implying $\omega<0$. 
On the   two-loop level,  we find a negative prefactor $X_2<0$ and, 
thus, still $\zeta_{{\rm FRG},\kappa}<\zeta_{th,\kappa}=3/2$.

In the literature, the existence of an upper 
 critical dimension $d_u$, above which
 $\zeta<\zeta_{th}$ applies, has been discussed, in particular for  DLs 
($D=1$, $m=1$) and 
as a candidate for the upper critical dimension of the KPZ equation to 
which the DL problem can be mapped.
Our above finding  $\zeta_{{\rm FRG},\kappa}<\zeta_{th,\kappa}=3/2$ in two-loop 
order indicates that for SDLs with $m=2$, $d=1$ is already above the upper 
critical  dimension $d_{u,\kappa}$, i.e., $d_{u,\kappa}<1$.
Using the one-loop result for general $d$,
we can estimate this upper critical dimension $d_{u,\kappa}$ for SDLs.
The approximate formula
\begin{equation}
   \zeta_{\rm FRG}(d) =  \frac{\varepsilon}{4+d}
  \left( 1+ \frac{1}{4e} 2^{-(d+2)/2}\frac{(d+2)^2}{d+4} \right)
\end{equation}
from Ref.\ \cite{Balents1993} gives a critical dimension
$d_{u,\kappa} = 0.937669<1$. 
Solving the FRG fixed point equation for $R({\bf z})$ 
numerically, we  find $d_{u,\kappa} \approx 0.84 <1$ using the
 one-loop equations and the numerical methods outlined in
\cite{LeDoussal2004}
 (we  use Taylor expansions up to order $12$, which allows us to
 determine $\zeta$ to four digits). The result $d_{u,\kappa}<1$ remains
valid using the two-loop equations, where the additional two-loop terms contain
the
same factor $X_m$ as for $d=1$. However, the two-loop result for the critical
dimension is lower because, analogously to the DL results of Ref.\
\cite{LeDoussal2005}, the two-loop corrections are negative for
$d>d_{u,\kappa}$.
In the following section, we present numerical results 
which  show that there exists 
a disorder dominated phase with $\zeta_{\kappa}>\zeta_{th,\kappa}=3/2$
for SDLs in $d=1$ below a critical temperature, 
which shows that the FRG results are questionable 
for lines with $D=1$ and, correspondingly,  large values for the 
expansion parameter $\epsilon$.

%%%%%%%%%%%%%%%%%%%%%%%%%%%%
\section{Numerical results} \label{sec:numericalresults}

Returning from (D+d)-dimensional manifolds 
to the problem of (1+d)-dimensional SDLs in disorder, 
further progress is possible by  numerical studies using the 
 transfer matrix method \cite{HalpinHealy1995,Wang2000} both 
for $T=0$ (see Fig.\ \ref{fig:t0}) and for  $T>0$. 

Previous numerical studies for DLs in
1+3 dimensions offer the opportunity for a comparison of the exponents
$\omega$ (energy fluctuations), describing the low temperature
phase, and $\nu$ (correlation length), describing the transition\footnote{To
fully describe the behavior at criticality an infinite number of exponents is
needed, see Refs.\ \cite{Mukherji1996,Monthus2007} and section 
\ref{multifrac}.}, 
in  the
low temperature phase to test the aforementioned analogy to a SDL in
1+1 dimensions. The values found there
are
$\omega \approx 0.186$ \cite{Kim1991b,Ala1993,Monthus2006a,Marinari2000}  and
either $\nu\approx 2$ \cite{Monthus2006a,Monthus2007,Derrida1990}
or  $\nu\approx 4$ \cite{Monthus2006a,Kim1991}.

%%%%%%%%%%%%%%%%%%%%%%
\subsection{Transfer matrix algorithm}

From a discretization of the energy functional given in eq.\ \ref{eq:ham}
we conclude that a segment of a SDL with length $\Delta L = 1$ starting at $z$
with orientation $\text{d}z/\text{d}x=v$ and ending at $z'$ with orientation
$v'$ contributes an energy 
\begin{align}
\Delta E_x(v-v',z')=\frac{\kappa}{2}(v-v')^2+V(x,z'), 
\end{align}
where the additional condition  $z'=z+v$ applies which connects
positions and tangents. 
We will also 
exploit the quadratic behavior in $v-v'$ and only consider  segments
with $(v-v')^2\leq \Delta_v^2$ with a constant $\Delta_v >1$ 
in the partition sum. 
The random potential $V(x,z)$ is represented by
Gaussian random variables \cite{Marsaglia2000} with $\overline{V(x,z)}=0$ and
$\overline{V(x,z)V(x',z')}=g^2\delta_{xx'}\delta_{zz'}$. For the sake of
simplicity and comparability we choose $g=1$ and vary the temperature. Unless
stated differently we set $\kappa=5$ and $\Delta_v=5$ and simulate lengths 
up to $L=100$ using, in principle, all intermediate lengths  
and at least $10^4$ realizations of the disorder.

%%%%%%%%%%%%%%%%%%%%%%
\subsubsection{Zero temperature}

For vanishing temperature $T=0$ there are no entropic contributions to the free
energy. Hence the line is always in its ground state and minimizing the
energy $E_L^0$, where the subscript denotes the length and the superscript is a
reminder that these considerations are valid for $T=0$.
This can be done iteratively
\begin{align}
 E_{L}^{0}(z',v')&{=}\min_{v} \left[E^{0}_{L-1}(z'{-}v,v)+\Delta
E_{L}(v{-}v',z') \right]
\end{align}
and all quantities $X$ are to be measured in the resulting 
(non-degenerate) ground state
\begin{align}
 \langle X \rangle_{T=0} &= X(z_\text{min},v_\text{min})
\end{align}
with $E^0_L(z_\text{min},v_\text{min})=\min_{z,v} E^0_L(z,v)$.  

%%%%%%%%%%%%%%%%%%%%%%%%%
\subsubsection{Finite temperature}

At finite temperatures $T>0$, the restricted partition function
$Z_{L}(z,v)=\int_{z(0)=v(0)=0}^{z(L)=z,v(L)=v} {\cal D}z
\exp{(-\beta{\cal H})}$ has to be evaluated. We are using the transfer matrix
approach and divide the line into straight segments which are connected by
the {\em transfer matrix}. This allows for an iterative computation of the
restricted partition function 
\begin{align}
 Z_{L}(z',v')&=\!\!\!\!\! \sum_{\substack{v,\\\!z=z'-v\!}}\!\! \!\!
\exp{({-}\beta\Delta E_{L}(v{-}v',z'))} Z_{L-1}(z,v) ; 
\label{eq:tm}
\end{align}
for numerical stability reasons, the $Z_{L}(z,v)$ are normalized 
 in each length iteration such that $\sum Z_L(z,v)=1$.
The normalization constant is a useful quantity as it is the total partition
function and therefore gives the free energy. Additionally, the normalized
restricted partition function is used in the computation of thermal averages
\begin{align}
 \langle X \rangle =  \left(\sum_{z,v}
Z_L(z,v)\right)^{-1} \sum_{z,v} X(z,v) Z_L(z,v)
\end{align}
This averaging procedure is only correct for quantities $X$ that are measured at
the end of the line, moments of the energy (potential or total) are
accumulated along the contour of the line and have, therefore, to be computed
in an iteration scheme \cite{Wang2000} very similar to eq.\ \eqref{eq:tm}. 
Finally, for
all observables the quenched average over realizations of the disorder 
has to be performed.

%%%%%%%%%%%%%%%%%%%%%
\subsection{Existence and nature of the disorder dominated phase}
\label{sec:existence}

%%%%%%%%%%%%
\subsubsection{Roughness}  \label{sec:rough}

\begin{figure}[t]
\includegraphics[width=0.45\textwidth]{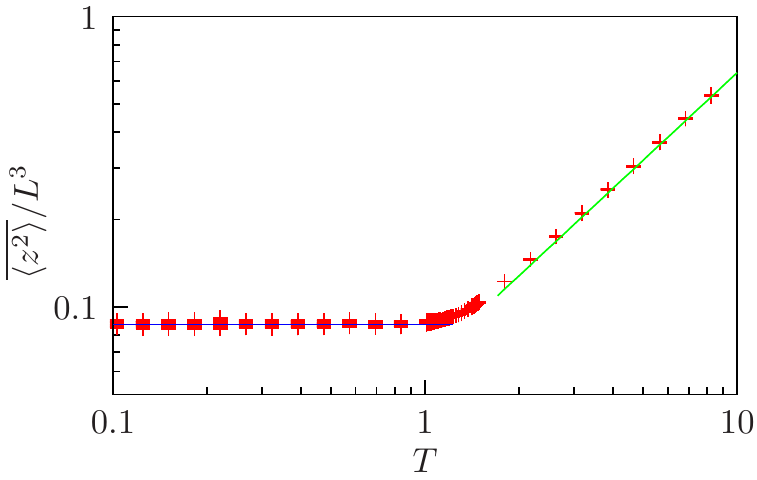}
\caption{
The roughness $\overline{\langle z^2\rangle}$ as a function of the
temperature. There are two distinct regimes where the roughness scales as
$\overline{\langle z^2\rangle}\sim T^0$ and $\overline{\langle z^2\rangle}\sim
T$ respectively. We plotted these asymptotics as well. We show  lengths
$L=50,60,\ldots,100$, so that the ``scattering'' of the symbols for one
temperature indicate the quality of the rescaling with $L^{-3}$ 
(see also Fig.\ \ref{fig:zeta}).
}
\label{fig:z2vsT}
\end{figure}

\begin{figure}[t]
\begin{center}
 \includegraphics[width=0.45\textwidth]{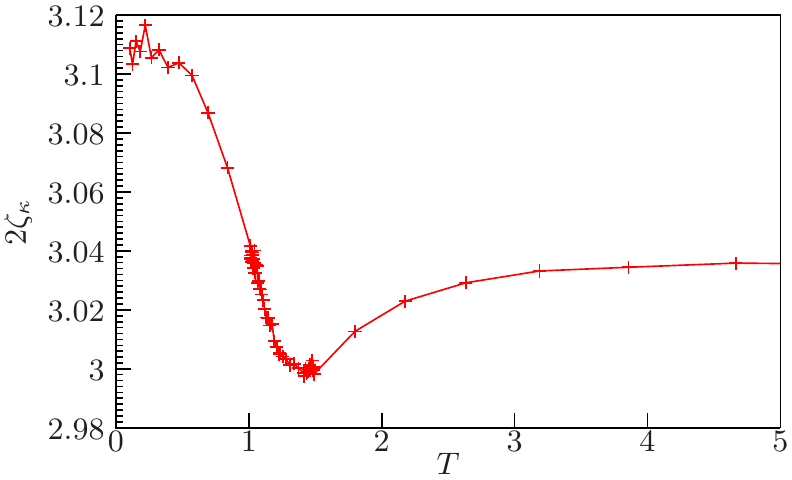}
 \caption{(color online)
   Roughness exponent $2\zeta_{\kappa}$ for various temperatures,
   computed via \eqref{eq:loczeta}. The deviation for high
   temperatures from the analytical value $2\zeta=3$ indicates
   numerical shortcomings, nonetheless there is a clear ``dip'' at
   $T\approx 1.4$, which we identify as the critical temperature. For
   low temperatures, we find values consistent with
   $\omega\approx 0.11$.}
\label{fig:zeta}
\end{center}
\end{figure}

The  most natural observable to analyze looking for a localization
transition is the roughness $\overline{\langle z^2\rangle}$. One expects to
see two different regimes, a high temperature phase with 
the thermal roughness $\overline{\langle
z^2\rangle}\sim T L^3$ and a low temperature phase with $\overline{\langle
z^2\rangle}\sim T^0 L^{2\zeta}$. Also, the influence of numerical details
onto the roughness  should be smaller  than their influence 
 on the free energy. 
Figure \ref{fig:z2vsT} shows the roughness $\overline{\langle z^2\rangle}$
as a function of temperature and demonstrates that these
expectations are met. In order to  determine the roughness 
exponent $\zeta_{\kappa}$ we 
measure a  ``local'' roughness exponent \cite{Schwartz2012}
\begin{align}
 2\zeta(L)=\log_5 (z^2(L)/z^2(L/5)) 
\label{eq:loczeta}
\end{align}
The data for $\zeta_\kappa$ as a function of temperature 
presented in Fig.\ \ref{fig:zeta} exhibits two distinct
 high and low temperature regimes  and a significant ``dip'' of the local
roughness exponent around $T\approx  1.4$.
 This method of determining $\zeta$ gives
better results than fitting $\overline{\langle z^2\rangle}(L)$
% at some $T\ll 1.4$. 

Via the scaling relation (\ref{eq:scalrel}), $\omega=2\zeta_{\kappa}-3$, 
we obtain the exponent $\omega$ from the roughness exponent $\zeta_\kappa$. 
As in the context of  DLs \cite{Doty1992}, it can be argued that $\omega$
should vanish at the disorder-induced localization 
transition, resulting in a roughness exponent
$\zeta_{\kappa} = 3/2$.  This seems to hold, even though the numerical
value for high temperatures is slightly above $\zeta_{\kappa}=3/2$.
 This is strong evidence for  a phase transition at
$T_c\approx 1.4$. For low temperatures, the 
values $2\zeta_{\kappa}\approx 3.11$
give    $\omega\approx 0.11$ according to eq.\ 
(\ref{eq:scalrel}), which is slightly lower than $\omega =0.186$, 
which is the literature value for DLs in 1+3 dimensions 
\cite{Kim1991b,Ala1993,Monthus2006a,Marinari2000}.

%%%%%%%%%%%%
\subsubsection{Disorder-induced persistence length}

The roughness is closely related to the 
averaged tangent directions
$\overline{\langle v^2\rangle}\equiv
\overline{\langle (\partial_x z)^2\rangle}$,
which should scale as  
\begin{equation}
  \overline{\langle v^2\rangle}\sim 
 \overline{\langle z^2\rangle}/L^2 \sim L^{2(\zeta-1)}\sim 
   L^{1+\omega}.
\end{equation} 
We define an effective disorder-induced persistence length $\tilde{L}_p$
for the SDL 
as the length scale at which the tangent fluctuations
become equal to one
\begin{equation}
 \overline{\langle v^2\rangle}(\tilde{L}_p) = 1 .
\label{eq:deflp}
\end{equation}
This generalized definition for the disordered system 
is consistent with the standard definition for the persistence 
length $L_p$ of a thermally fluctuating SDL in the absence of 
disorder,  where we expect
$\overline{\langle v^2\rangle}\approx L/L_p$ with
$L_p=  \beta \kappa$.
Apart from numerical prefactors this gives the 
standard persistence length of the  WLC model, see eq.\ \eqref{eq:lp},
which is defined as the decay length of tangent correlations. 

In the low
temperature phase, we expect a disorder dominated roughness and, therefore,
 temperature-independent tangent fluctuations $\overline{\langle v^2\rangle}$,
which results in a temperature-independent  {\em
disorder-induced persistence length} $\tilde{L}_p \sim \beta^0$.
 The line roughens in the low
temperature phase, which 
gives rise to  a persistence length {\em decrease} as compared to the 
thermal persistence length $L_p\sim \kappa/k_BT$. 
From  the Flory-result $z\sim(g/\kappa)^{2/(4+d)}  L^{7/(4+d)}$ at 
low temperatures, we expect $v\sim (g/\kappa)^{2/(4+d)}  L^{(3-d)/(4+d)} $ 
for the disorder-induced tangent fluctuations and 
\begin{equation}
  \tilde{L}_p \sim (\kappa/g)^{2/(3-d)}
\label{eq:lpFl}
\end{equation}
at low temperatures according to the criterion (\ref{eq:deflp}). 
For SDLs in 1+1 dimension, we expect $\tilde{L}_p \propto g^{-1}$. 

\begin{figure}[t]
\includegraphics[width=0.45\textwidth]{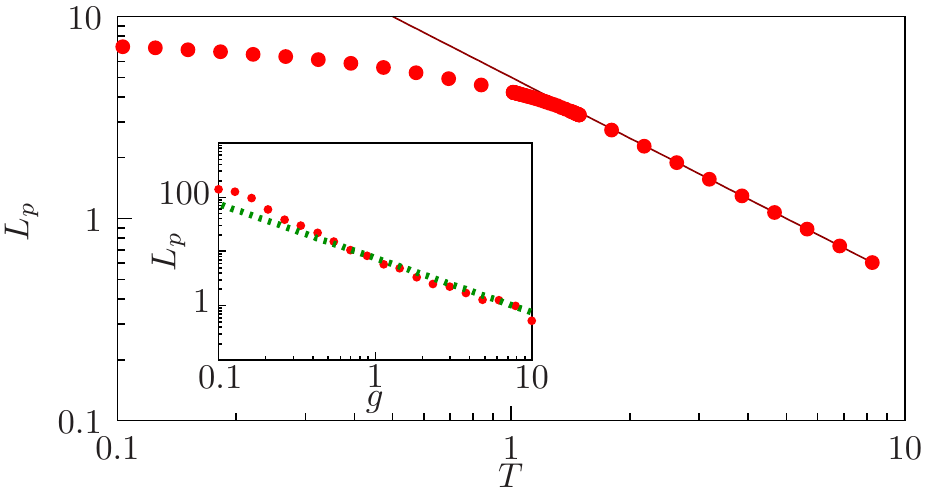}
\caption{(color online) 
The generalized  persistence length $\tilde{L}_p$ according to eqs.\ 
\eqref{eq:deflp}  and \eqref{eq:fitv22} as a function of the temperature $T$. 
\eqref{eq:fitv22}. $\tilde{L}_p$ matches its thermal
value (solid line) for $T>T_c$ and is reduced and approximately constant for
$T<T_c$. Inset:  $\tilde{L}_p$ at $T=0$ versus the potential strength
$g$. The Flory-result $\tilde{L}_p\sim g^{-1}$ (dotted line), see eq.\  
(\ref{eq:lpFl})
matches the data.
}
\label{fig:lp3}
\end{figure}

To determine the generalized persistence length from 
eq.\ (\ref{eq:deflp}) numerically, we used the 
 {\em tilt symmetry} of the replicated 
Hamiltonian \cite{Giamarchi1995}, according to which the ``connected''
average
\begin{equation}
 \overline{\langle v^2\rangle - \langle v\rangle^2} \approx L/\beta \kappa
\end{equation}
is independent of  the disorder strength. Therefore, we can use a fit to the
sample-to-sample tangent fluctuations of the form
\begin{equation}
 \overline{\langle v\rangle^2}(T,L)=a(T)L^{1+\omega'(T)} 
\label{eq:v2fit}
\end{equation}
with an amplitude   $a(T)$ and an exponent $\omega'(T)$, which 
should agree with the free energy exponent $\omega$.  
Then, we can rewrite $\overline{\langle v^2\rangle}$ as
\begin{align}
\overline{\langle v^2\rangle}&=
 \overline{\langle v^2\rangle - \langle v\rangle^2}
     +\overline{\langle v\rangle^2}
\nonumber\\
&=L/\beta\kappa+a(T)L^{1+\omega'(T)},
\label{eq:fitv22}
\end{align}
and determine $\tilde{L}_p$ using  eq.\ \eqref{eq:deflp}.
Our results for $\tilde{L}_p$ are shown in Fig.\ \ref{fig:lp3}. 
In the low-temperature phase for $T<T_c\approx 1.4$, the 
 persistence length becomes  indeed disorder-induced, i.e.,  almost 
independent of  temperature and our results are consistent 
with the above scaling result (\ref{eq:lpFl}).  We consider this
non-divergent persistence length at low temperatures also 
to be consistent with previous 
results for the non-directed version of the SDL, the WLC in
disorder \cite{Dua2004}. 
In the high temperature phase, our results approach  the
standard  thermal persistence length 
$L_p \sim \beta \kappa$. 

\begin{figure}[t]
 \includegraphics[width=0.45\textwidth]{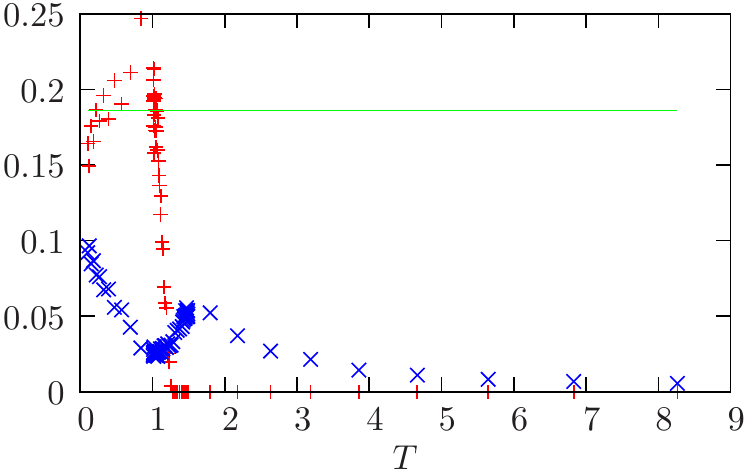}
 \caption{(color online) 
Parameters $a(T)$ (blue, Greek crosses) and $\omega'(T)$ (red, saltires)
used in
fitting $\overline{\langle v\rangle^2}$, cf.\ eq.\ \ref{eq:v2fit}. 
The latter vanishes around, but not exactly at $T=T_c$. 
The horizontal line corresponds to the literature 
value for DLs in 1+3 dimensions, $\omega=0.186$. 
}
 \label{fig:abomega}
\end{figure}

The fit results for 
$a(T)$ and $\omega'(T)$ are shown
in Fig.\ \ref{fig:abomega}. For $T<T_c$ our results are consistent 
with $\omega\approx 0.186$, which is the 
literature value for DLs in 1+3 dimensions 
\cite{Kim1991b,Ala1993,Monthus2006a,Marinari2000}. 
In fact, we get very similar results for
$L_p$  if we fix $\omega'(T)=\omega=0.186$.

%%%%%%%%%%%%%%%
\subsubsection{Derrida-Flyvbjerg singularities}

\begin{figure}[t]
\includegraphics[width=0.45\textwidth]{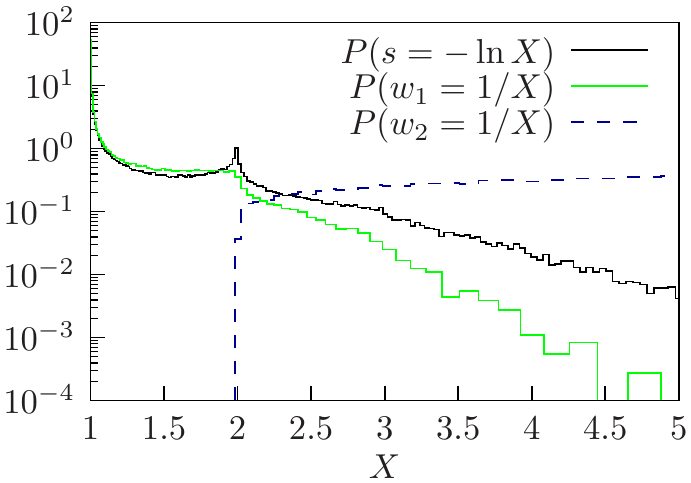}
\caption{(color online) 
Distribution of the largest ($w_1$, light (green) solid line) and second
largest
($w_2$, (blue) dashed line) statistical weight $w(z)=Z(z)/Z$ and the 
information entropy $s=-\sum_z
w(z)\ln{w(z)}$ (black, dark solid line).} 
\label{fig:derrflyv}
\end{figure}

One of the features expected in a disorder dominated phase are 
Derrida-Flyvbjerg singularities \cite{Derrida1987,Monthus2007b}. These are
features of  the statistical weights, here the normalized restricted
partition $w(z)=Z(z)/Z$ function for the SDL to end at a specific 
point $z$. As the
phase is disorder dominated, some characteristics of these weights should origin
from mere statistics and the distribution used for the random potential rather
than details of the Hamiltonian involved. 
According to Derrida and Flyvbjerg the distributions $P_1(w_1)$ and
 $P_2(w_2)$ 
of the  largest and the second largest weight, respectively, 
 should exhibit singularities at $1/n$ ($n=1,2,3,...$) in a 
disorder dominated phase with a multivalley structure of phase space 
\cite{Derrida1987}.
For SDLs in disorder, we calculated the  distribution of the
value of the largest and the second largest weight numerically 
as shown in Fig.\ \ref{fig:derrflyv} and indeed find 
singularities at $1$ and $1/2$ for  $T<T_c$. 
We are not able to clearly resolve
higher singularities at $1/n$ with $n \ge 3$, which might be due to
the underlying (Gaussian) distribution or the number of samples used. 
Analogous
singularities can be found in the distribution of the information entropy
$s=-\sum_z w \ln w$ at values $-\ln(1),-\ln(2),-\ln(3)$, whereas nothing
similar can be observed at high temperatures, where the entropy distribution is
Gaussian and the distribution of the (second) largest weight is sharply peaked
around zero. We see
this as a confirmation that the SDL indeed undergoes a transition 
to a disorder dominated phase At $T_c \approx 1.4$.

%%%%%%%%%%%%
\subsection{The free energy distributions of 
  SDL in 1+1 and DL in 1+3 dimensions are identical} \label{sec:distrib}

\begin{figure}[t]
\begin{center}
 \includegraphics[width=0.45\textwidth]{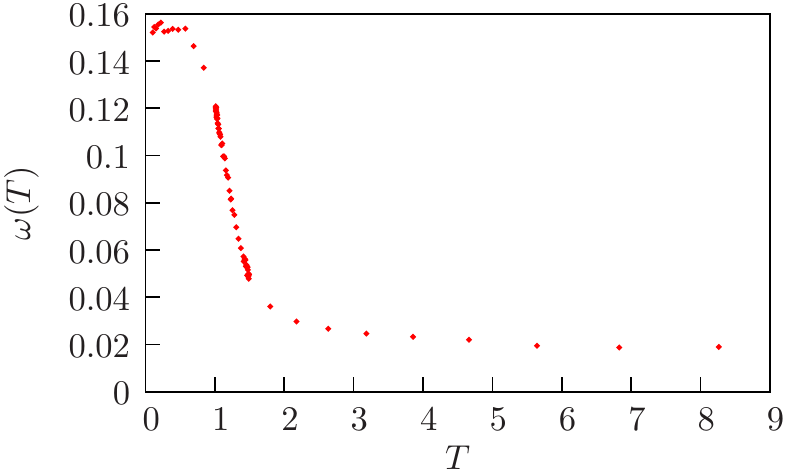}
 \caption{(color online) Fluctuation exponent $\omega$ determined by the free
energy fluctuations.}
\label{fig:omegadeltaf2}
\end{center}
\end{figure}

\begin{figure}[t]
\begin{center}
 \includegraphics[width=0.45\textwidth]{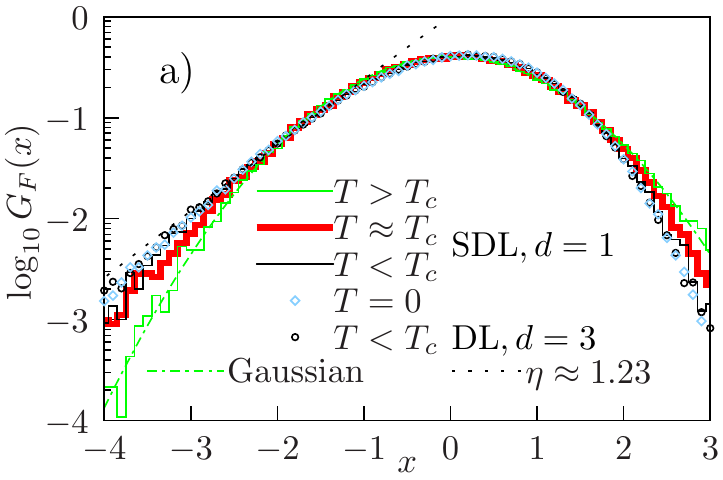}
 \caption{(color online)
   Rescaled $G_F(X)=P((F-\overline{F})/\Delta F)$
free energy distribution for a stiff directed line (SDL) in 1+1
dimensions. 
  We show distributions for $T=0$ (light (blue) squares, ground state energy) as
well as
for three finite  temperatures $T<T_c$ (black thin solid line), $T\approx T_c$
(red, thicker solid line) and
   $T>T_c$ (green, light thin solid line). Results for a directed line (DL) 
  in 1+3 dimensions are shown (dark circles).}
\label{fig:g_f}
\end{center}
\end{figure}

The exponent $\omega$ can alternatively be determined in a more direct way by
fitting  $\Delta F=({\overline{F^2}-\overline{F}{}^2})^{1/2}\propto L^\omega$
at temperatures $T\ll T_c$, giving values of 
$\omega\approx 0.15-0.16$ (cf.\ Fig. \ref{fig:omegadeltaf2}).

We can go one step further and consider not only the second moment 
but the whole  distribution of the free
energy as shown in Fig.\ \ref{fig:g_f}, which is obtained  by computing the
free energy for every sample and  rescaling to zero mean and
unit variance,
\begin{align}
G_F(X)=\text{Prob}((F-\overline{F})/\Delta F=X).
\end{align}
This rescaling should make $G_F$ more robust against the influence of
numerical details. For DLs in $d=1$ it has been found that this
distribution is of a universal form \cite{Dotsenko2011}. 
The asymptotic behavior
of the negative tail of the rescaled
free energy distribution for low temperatures is of the form
\begin{align}
\ln{G_F(X)}&\sim -\lvert X\rvert^\eta \quad (X<0,\lvert X\rvert\gtrapprox 1).
\end{align}
This allows us to  determine  the energy fluctuation exponent $\omega$
via the Zhang argument \cite{HalpinHealy1995,Monthus2006a,Monthus2006c}: 
a saddle point integration gives $\ln \overline{Z^n}\sim -n \overline{F}/T 
   -(n \Delta F/T)^{\eta/(\eta-1)}$; on the other hand, $\ln \overline{Z^n}\sim
L$ should be extensive resulting in $\Delta F \sim L^{1-1/\eta}$ or 
\begin{align}
\eta=(1-\omega)^{-1}. \label{eq:eta}
\end{align}
 We find  $\eta\approx1.23$ (dashed black line in Fig.\ \ref{fig:g_f}) or
$\omega\approx 0.18$.  This is in agreement with the values reported for
DLs in 1+3 dimensions.

\begin{figure}[t]
\includegraphics[width=0.45\textwidth]{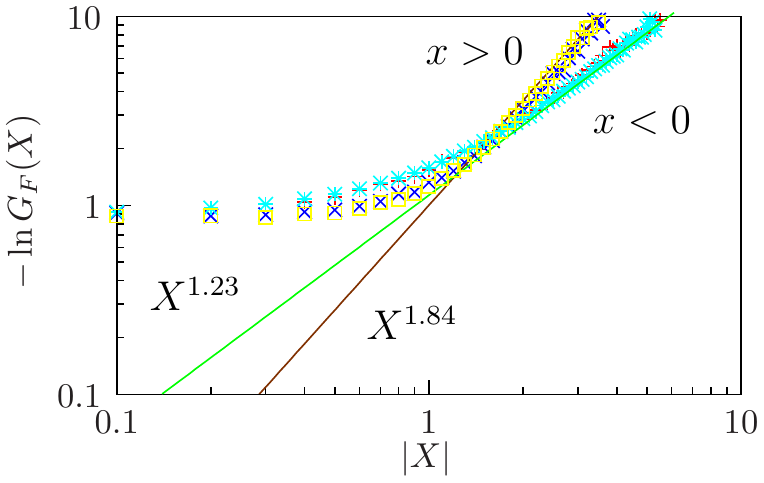}
\caption{(color online) Double logarithmic plot of the negative
logarithm of
the rescaled free energy distribution $G_F(X)$ ($X=(F-\overline{F})/\Delta F$)
at low temperatures $T<T_c$ for
the DL (light blue stars for $X<0$ and yellow squares for $X>0$) and the SDL
(red Greek crosses for $X<0$
and dark blue saltires for $X>0$). We see identical  behavior for SDL and DL
 consistent
with exponents  $\eta\approx 1.23$ and $\eta'\approx 1.84$, see 
text.
}\label{fig:eta}
\end{figure}

For a direct comparison of the rescaled free energy
distributions of a SDL in 1+1  and a   DL in 1+3 dimensions we simulated
both systems (the DL up to lengths $L=60$) and find  that the rescaled free
energy distributions in the low temperature phases have to be considered {\em
identical} within numerical accuracy. This could hint towards a
new seemingly ``universal'' distribution for certain random systems, much like
the Tracy-Widom distribution that is found for the DL in 1+1 dimensions and
various other systems \cite{Dotsenko2011}. 
For finite system sizes this universal
behavior can only be expected for free energy fluctuations 
 $\lvert X\rvert$ small compared to
an upper threshold \cite{Kolokolov2007,Kolokolov2008}; we believe,
 however, that our simulation does not cover 
the very rare fluctuations that induce the
non-universal part for very large $\lvert X\rvert$. 
In Fig.\ \ref{fig:eta} we
show the distributions for $T<T_c$ (DL and SDL) in a manner where the exponent
$\eta$ becomes more apparent. Here we introduce the exponent $\eta'$, which is
the analogon to $\eta$ for the positive tail
\begin{align}
\ln{G_F(X)}&\sim -\lvert X\rvert^{\eta'} \quad (X>0,\lvert X\vert\gtrapprox1),
\end{align}
where the Zhang argument is not applicable. 
We find a value $\eta'\approx  1.84$.

Based on an exact renormalization on the diamond
lattice \cite{Monthus2008} and an optimal fluctuation
approach \cite{Kolokolov2008}, it has been previously suggested
for DLs  that $\eta'$ and $\omega$ 
are related via (cf.\ eq.\  \eqref{eq:eta})
\begin{align}
 \eta'&=d_{\text{eff}}/(1-\omega)
\end{align}
with $d_{\text{eff}}=1+d$ for the hypercubic lattice.  
This is found 
to be valid for the DL in 1+1
dimensions, where $\omega=1/3$ and $\eta'=3$
 \cite[and references cited therein]{Monthus2008}. For the problem at hand,
the literature value  
$\omega=0.186$ for DLs in  $d=3$ would lead to $\eta'\approx 5$, which is
far from the value  $\eta'\approx  1.84$ we find;
 thus, the ratio $\eta'/\eta$ does not match the
prediction $\eta'/\eta=(1+d)$. In Ref.\ \cite{Kolokolov2008} $\eta=2$ and
$\eta'=3$ were found to be superuniversal (independent of $d$) for the DL in
dimensions $d>2=d_c$ at temperatures $T>T_c$. We can neither confirm nor deny
this result, as we are not able to cover the ``most distant'' part of the
distribution, but our interpretation that $G_F$ is Gaussian for $T>T_c$ would
lead to $\eta=\eta'=2$.

The finite size corrections to the free
energy\cite{HalpinHealy1995,Krug1990}
should also scale as $L^\omega$ leading to
\begin{align}
 F/L &\approx a + b L^{\omega-1} 
\end{align}
We did not succeed in determining a precise, consistent $\omega$ in this
way, possibly
because our systems are too small.

\begin{figure}[t]
\includegraphics[width=0.45\textwidth]{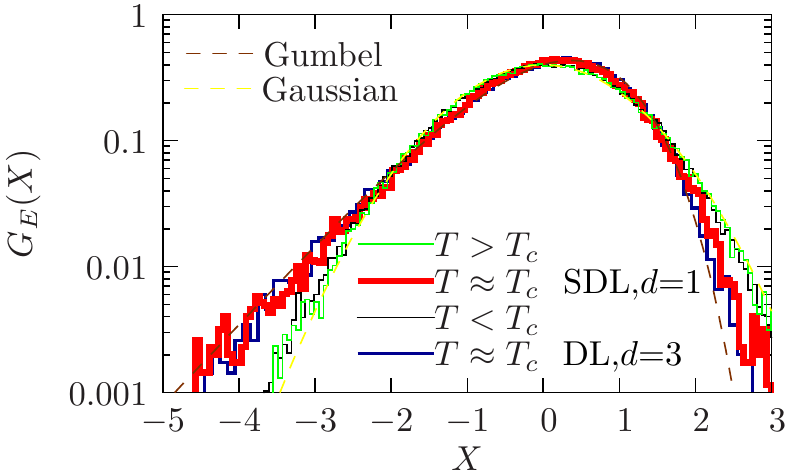}
\caption{ 
   Rescaled potential  energy
   distribution for a SDL in 1+1 dimensions.
Plotted are distributions for three
   finite temperatures  $T<T_c$ (black, dark thin solid line), $T\approx T_c$
(red, thick solid line) and
   $T>T_c$ (green, light solid line). Results for a DL in 1+3 dimensions are
shown 
   in blue (medium solid line). 
Here, the brown (dark dashed) curve is an approximation using a one-parameter
Gumbel
distribution (cf.\ eq.\ \eqref{eq:gum} with $m=1.7$), 
whereas the yellow (light dashed) curve is
the normal distribution.}
\label{fig:g_e}
\end{figure}

The free energy distributions 
 in Fig.\ \ref{fig:g_f}  seem to
be identical for $T<T_c$ and $T\approx T_c$ 
(as it is also the case  for the DL \cite{Monthus2006a}),
 which suggests
 $\omega_{T=T_c}=\omega_{T<T_c}\approx0.186$ contradicting the
$\omega_{T=T_c}=0$ argument \cite{Doty1992}, but one has to bear in mind that
the saddle point integration in
the course of the Zhang argument is only applicable
 if $\eta>1$ or $\omega>0$.

 A more distinct difference between the behavior at criticality and at
low temperatures can be found in the distribution of the potential energy
as shown in Fig.\ \ref{fig:g_e}
(also the potential energy is rescaled 
 using $X=(E-\overline{E})/\Delta E$ analogously 
to Fig.\ \ref{fig:g_f} for the free energy distribution). 
For the potential energy distribution, 
the behavior at the critical temperature is clearly
different from the behavior both at 
$T>T_c$ and $T<T_c$ (which are not identical for the
DL) and exhibits a   decay 
\begin{align}
 \ln{G_E(X)}\sim\exp{(-|X|)} \quad (X<0,\lvert X\vert\gg1).
\end{align}
resembling extreme value distributions of the Gumbel type. 

A tentative explanation might be that the transition occurs because of
extreme values of the potential at which the otherwise thermally fluctuating
line localizes. The random potential has a Gaussian distribution, thus its
extreme values are distributed according to a Gumbel distribution or
Fisher-Tipett type I extreme value distribution \cite{Gumbel1954}
\begin{align}
 P^{\alpha,\beta}_{\text{Gumbel}}(X)&=\frac{1}{\beta}e^{-z(X)-e^{-z(X)}},
\end{align}
with $z(X)=(X-\alpha)/\beta$, the location parameter $\alpha$, and the scale
parameter $\beta$ (the parameters depend on the number of ``trials''). 
The distribution of the potential energy at the critical 
temperature might be of
a similar shape.
As we are studying the rescaled distribution of the
potential energy, which has zero mean and unit variance, we can use a
one-parameter version of the Gumbel distribution
\begin{subequations}
\begin{align}
 g_m(X)&=\frac{\psi_1(m)m^m}{\Gamma(m)}
      \exp{\left(h_m(X)-\mathrm{e}^{h_m(X) } \right) }^m 
\label{eq:gum}
\end{align}
with the {\em shape parameter} $m$, the abbreviation
\begin{align}
h_m(X)&\equiv (X+\psi(m))\psi_1(m)-\ln{m}
\end{align}
\end{subequations}
and the usual 
$\Gamma(m)=\int_0^\infty\mathrm{d}t\,t^{m-1}\mathrm{e}^{-t}$ (gamma function),
$\psi(m)=\mathrm{d}\ln{\Gamma(m)}/\mathrm{d}m$ (digamma function) and
$\psi_1(m)=\mathrm{d}^2\ln{\Gamma(m)}/\mathrm{d}m^2$ (trigamma function)
\cite{Abramowitz1972}.
This distribution inherently features the correct decay at the tails, with
$\ln{g_m(X)}$ decreasing faster than algebraic for $X<0$ and 
linearly with slope
$m\psi_1(m)$ for $X>0$. However, our best approximation to the data with 
$m=1.7$ deviates for $X>0$.
It is remarkable  that the potential energy 
distribution is well described by an  extreme value 
distribution only right at the transition at $T=T_c$, whereas 
it approaches a Gaussian not only  for $T>T_c$ but also for 
$T<T_c$, see  Fig.\ \ref{fig:g_e}. The Gaussian distribution at low
temperatures might stem from the Gaussian distribution used in the realization
of the disorder potential.

\begin{figure}[t]
\begin{center}
 \includegraphics[width=0.45\textwidth]{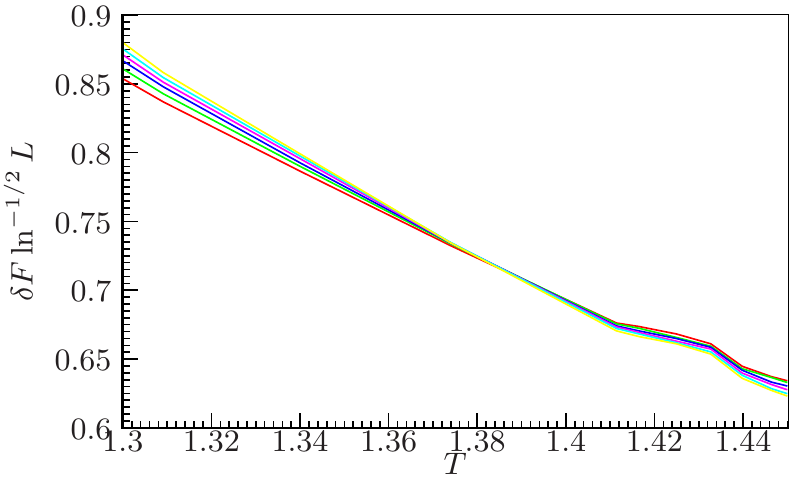}
 \caption{(color online)
   The reduced free energy 
    $\delta F=\overline{F}-\overline{F}_\text{ann}$
    rescaled by $\ln^{1/2}{L}$  for lengths
   $L=50,60,\ldots,100$ as a function of the temperature $T$. There is
   a pseudo-crossing of the lines around
   $T\approx1.38$. }
\label{fig:deltaf}
\end{center}
\end{figure}

At the critical temperature, where $\omega \approx 0$, 
the fluctuations of the free energy have been found to
scale logarithmically with $L$,  $\Delta F\sim \ln^{1/2}{L}$
\cite{Monthus2006a}.  We support this statement by studying the
difference of quenched and annealed free energies
$\delta F=\overline{F}-\overline{F}_\text{ann}$,
which should as well scale as $\delta F\sim\ln^{1/2}{L}$
at the transition. As the annealed free energy  directly depends
numerical details such as system size and neglected elements of the transfer
matrix, we determine it by simulating a
system without disorder and adding the contribution of the annealed
potential, $\overline{F}_\text{ann} = F_{g=0}-Lg^2\beta/2$.
This also allows for an approximate determination of  the critical
temperature by finite size scaling, see Fig.\  \ref{fig:deltaf}. 
To verify this
argument we tried also different scaling behaviors of the form 
$\delta F\sim\ln^c{L}$ and obtained a  consistent pseudo-crossing only
for $c\approx 1/2$. 
Furthermore the annealed free energy can be
used to compute the temperature $T_{0,\kappa}$, below which the annealed entropy
is negative. This provides a 
 lower bound on the actual critical temperature, whereas
the replica pair binding temperature 
$T_{2,\kappa}$ represents  an upper bound \cite{Monthus2006a}. We find
$T_{0,\kappa}\approx 0.4$, which is consistent with $T_c\approx 1.4$.

%%%%%%%%%%%%%%%%%%%%%%%%%%%%
\subsection{The localization transition temperature $T_c$ equals 
  the replica pair transition temperature $T_2$ for SDLs} 
\label{TcT2}

\begin{figure}[t]
\begin{center}
 \includegraphics[width=0.45\textwidth]{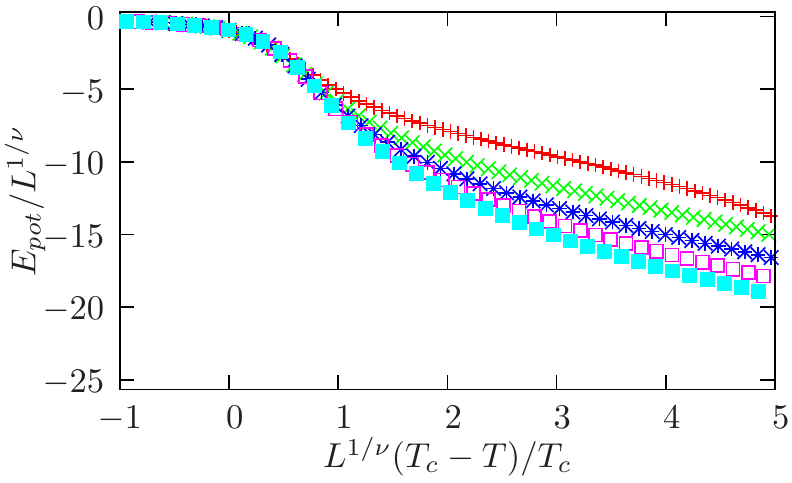}
 \caption{(color online) Finite size scaling for the SDL 
adsorption problem that
corresponds to the two-replica binding, see text for detailed
explanation. The scaling uses $T_c=1.44$ and $\nu=2$. 
As only one sample is needed
for the calculation we used larger system lengths
$L=100$ (red, Greek crosses),$200$ (green, saltires),\ldots,$500$ (light
blue, full squares).}
\label{fig:finscalt2}
\end{center}
\end{figure}

We have checked for the SDL via numerical  transfer matrix
calculations that the localization temperature in disorder $T_{c,\kappa}$
equals transition temperature $T_{2,\kappa}$ for 
 replica pair binding, $T_{c,\kappa}=T_{2,\kappa}$. 
As stated before, we use  the same transfer matrix algorithm for the 
replica pair system with a 
short-range binding potential by exploiting 
that the binding of two SDLs can be rewritten as a binding problem of one
effective SDL in an external potential using relative
coordinates \cite{Kierfeld2003}. This SDL
has a bending stiffness of $\kappa'=\kappa/2$ and the ``potential energy'' we
are interested in (cf.\ eq.\ \eqref{eq:H_rep}) for the replica pair binding 
 is $-\beta g^2\int \text{d}x \delta(z)$. For the interpretation of
the simulation results one has to keep in mind that the energy functional is
temperature dependent and, therefore, derivatives of the 
free energy with respect
to the inverse temperature $\beta$ are not given directly by cumulants of the
internal energy. Using $E_b=\int\text{d}x(\partial^2_x z)^2$ and
$V=-g^2\int\text{d}x\delta(z)$ the partition function is given by
$Z=\int\mathcal{D}z\exp{(-\beta E_b - \beta^2 V)}$ and the free energy by
$F=-\beta^{-1}\ln Z$ implying that
\begin{align}
 \frac{\partial(\beta F)}{\partial \beta} &= \langle E_b + 2\beta V\rangle =
U + \beta \langle V \rangle
\end{align}
where $U$ is the total 
internal energy (treating $\beta V$ as a potential). Thus,
the derivative  of the difference of the
free energies with and without the adsorption term 
with respect to the inverse temperature, which   should give the
divergent correlation length $\beta\delta F = F_V -F_0\sim L(T_c-T)^\nu$, is
identical to the ``potential energy'' $E_{pot}=-\beta V$ and usual finite size
scaling should be applicable, cf.\ Fig.\ \ref{fig:finscalt2}. 
We retain the known
correlation length exponent $\nu=2$ for the adsorption
problem \cite{Kierfeld2003}. 
We see matching
curves for the used system sizes $L=100, 200\ldots500$ 
around $T-T_c \approx 0$ for $T_{2,\kappa} =1.44$, which equals 
$T_{c,\kappa}\approx 1.4$  for the SDL in disorder. 

%%%%%%%%%%%%%%%%%%%
\subsection{Pair overlap as order parameter}
\label{sec:overlap}

\begin{figure}[t]
\begin{center}
 \includegraphics[width=0.45\textwidth]{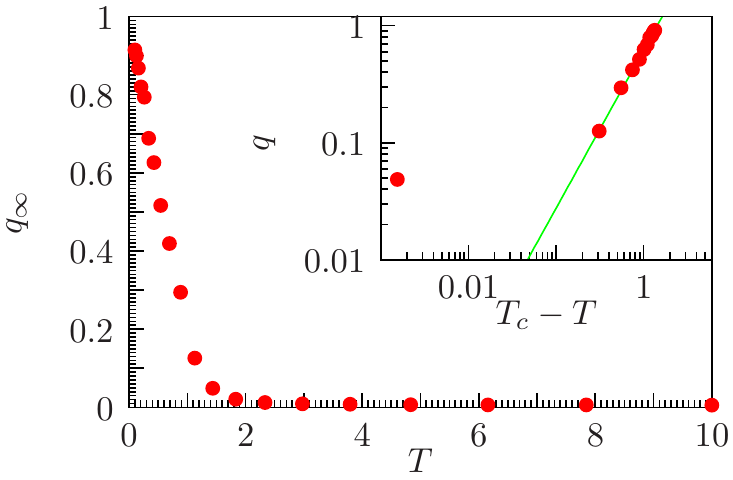}
 \caption{(color online) Overlap order parameter  $q$, as a
  function of $T$. We estimate $q$ from
 finite lengths using a fit $q_T(L)=a(T)/L+q_\infty(T)$.
 %, which is possibly not accurate for $T=T_c$ \cite{Mukherji1994}. 
Inset: Double-logarithmic plot
of the overlap $q$ versus  $T_c-T$ (with $T_c=1.44$), the
solid line is given by $q\sim (T_c-T)^{-\beta'}$ with $\beta'\approx-
1.36$.}\label{fig:overlap}
\end{center}
\end{figure}

\begin{figure}[t]
 \begin{center}
  \includegraphics[width=0.45\textwidth]{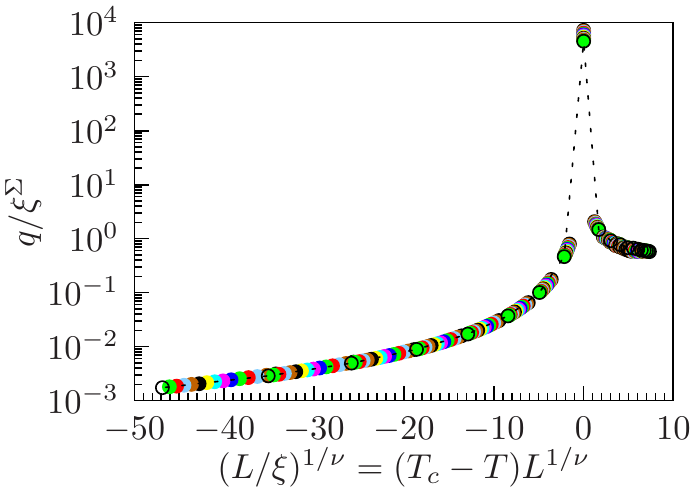}
 \caption{(color online) Finite size scaling ($L=15,\ldots,30$) for the overlap
order parameter. Our best results are $\Sigma\approx 0.75$ 
for the exponent related to the decay of the order
parameter, and $\nu\approx 2$ for the correlation length exponent. As before, we
used $T_c=1.44$. }\label{fig:overlap_fs}
 \end{center}

\end{figure}

Finally, we identify an order parameter of the localization
transition. 
For DLs, the disorder-averaged overlap 
$q= \lim_{L\rightarrow\infty}\overline{\frac{1}{L}\int_0^L\mathrm{d}x
\delta(z_1(x)-z_2(x))}$ of two replicas  has been proposed as order parameter
\cite{Mezard1990,Mukherji1994}. Up to now, it has been numerically impossible 
 to verify
this order parameter for DLs in $d>2$ dimensions where a localization 
transition exists because the relevant $2d$-dimensional 
two replica phase space is too large. 
For SDLs, on the other hand, 
   the transition is numerically accessible already in 
1+1 dimensions and we show that the overlap $q$ 
is indeed a valid order parameter using an adaptation
of the transfer matrix technique from Ref.\ \cite{Mezard1990}, see
Fig. \ref{fig:overlap}. This involves simulating {\em two} interacting SDLs,
therefore we can only use lengths up to $L=30$ and $10^3$ samples. 
For DLs, it has been
found  that the overlap at criticality decays as
$q\sim L^\Sigma$ with $\Sigma=-2\zeta=-(1+\omega)$ in $d=3$ 
\cite{Mukherji1994}. This has been
extended to
finite temperatures yielding $q\sim \lvert T-T_c\rvert^{-\nu\Sigma}$. Indeed, 
we find a
qualitatively similar behavior $q\sim \lvert T-T_c\rvert^{-\beta'}$ with 
 an exponent $\beta'\approx 1.3-1.4$.  Our best
estimate for $\Sigma$ 
 is $\Sigma\approx -0.75$, cf.\ Fig.\ \ref{fig:overlap_fs}. Because of  small 
simulation lengths  $L$ 
we do not conclude this deviation to be a definite statement against the
renormalization group results presented in Ref.\ \cite{Mukherji1994}, but
this would, unlike the DL renormalization group result $\Sigma=-(1+\omega)$,
suggest that two SDLs in a random potential are certain to meet
eventually \cite{Fisher1984}.
 For the correlation length exponent $\nu$ we find
values  $\nu\approx 2$ compatible with 
 the corresponding problem of DLs \cite{Monthus2006a,Monthus2007,Derrida1990};
such that our present results   deviate  
from $\beta'=\nu \Sigma$. Nevertheless, the connection between DLs and SDLs
provides the first system to test the proposed order parameter in a localization
transition numerically  and to determine the otherwise inaccessible exponents 
$\beta'$ or $\Sigma$.

%%%%%%%%%%%%%%%%%%%%%%%%%%%%%%%%%%%%%%
\subsection{Multifractal properties at the transition}
\label{multifrac}

For DLs in 1+3 dimensions, some insight into the 
underlying structure of the transition has been
achieved within the multifractal
formalism \cite{Jensen1987}. As we conjecture the transitions for the
DL and SDL to be essentially analogous, we expect to find similarities to the
analysis  that has been done before for DLs \cite{Monthus2007}, but also
deviations where the obvious geometrical differences become important. 
The idea
is that different moments $Y_q$ of the statistical weights
\begin{equation}
 Y_q(L) = \sum_{i} \text{Prob}(z(L)=z_i)^q
\end{equation}
are dominated by different regions of the support and thus show a different
scaling with the system size $L$. The
probability is given by the ratio of restricted and unrestricted partition
function $\text{Prob}(z(L)=z_i)\equiv w_L(z_i)=Z_L(z_i)/Z_L$. The sum goes over
all the possible ending points in the numerical simulation, whereas for the
continuous analytical problem $Y_q$ would be defined by an integral over $z$.

In the high temperature phase, the weights $w_L(z)$ obey the 
scaling form  $w_L(z)=L^{-\chi} \Omega(z/L^\zeta)$
with the return exponent $\chi$ (defined by 
$w_L(0)\sim L^{-\chi}$  as introduced above). 
It is straightforward to see that the
$Y_q$ then scale  according to 
\begin{equation}
\left. Y_q(L)\right|_{T>T_c} \sim L^{-(q-1)\chi}= L^{-(q-1)\zeta d}
\end{equation}
in the high temperature phase.
In the low temperature phase, the line is localized and therefore the $Y_q$
remain finite as $L\rightarrow\infty$. At criticality the $Y_q$ are
diverging, but the quenched disorder is relevant, and it becomes
important how the (necessary) average over realizations of the disorder is
computed. A common question regarding systems with disorder 
is whether a certain
quantity is self-averaging. If so its typical and average values, the latter of
which could be dominated by rare events, should be identical. For the
$Y_q$ a reasoning like this motivates the definition of
\begin{subequations}\label{eq:gendim}
\begin{align}
 Y_q^{av}=\left.\overline{Y_q}\right|_{T\approx T_c} &\sim L^{-\tilde{\tau}(q)}
= L^{-(q-1)\tilde{D}(q)}\\
Y_q^{typ}=\left.\exp{\overline{\ln{Y_q}}}\right|_{T\approx T_c} &\sim
L^{-\tau(q)}=L^{-(q-1)D(q)}\text{,} \label{eq:yqtyp}
\end{align}
\end{subequations}
where the definition of $D(q)$ is such that it includes the
obvious case of $q=1$, where $Y_1=1$ and $\tau(1)=0$ by definition. The $D(q)$
are referred to as
generalized dimensions \cite{Jensen1987}, and the function $D(q)$ 
discriminates between monofractal ($D(q)=const$ as for $T>T_c$) 
and multifractal behavior
($D(q)\neq const$). The interpretation of the $D(q)$ as (fractal) 
dimensions of
subsets is rather peculiar as we are dealing with a probability
distribution or {\em measure}. Nonetheless it is useful as it requires $D(q)$
to be monotonically decreasing because none of the subsets 
can have a larger dimension than their union. 
There are at least two  special
values of $D(q)$ with an obvious  meaning: $D(0)$ is the Hausdorff
dimension of the support and thus directly
related to the geometry of the system \footnote{The
Hausdorff dimensions for the measures related to DLs and SDLs 
with $d$ transverse dimensions are $D_{\tau}(0)=d$ and $D_{\kappa}(0)=2d$,
 ($z$ and $v=\partial_x z$), respectively,
which gives  $D_{\tau}(0)=3$ for DLs in 1+3 dimensions and 
$D_{\kappa}(0)=2$ for SDLs in 1+1 dimensions.}
and $D(1)$ is called the information dimension 
as it appears like a dimension in
the Shannon information entropy
\begin{align}
 s&=-\sum_i w(z_i)\ln w(z_i) 
 = - \left.\partial_q Y_q\right|_{q=1}\approx D(1)\ln L .
\end{align}
However $D(1)$ cannot be computed directly ($Y_1\equiv 1$) but only as an
analytic continuation
of $D(q)$. A measure is called fractal iff $D(0)>D(1)$ \cite{Tel1988}.

\begin{figure}[t]
\begin{center}
 \includegraphics[width=0.45\textwidth]{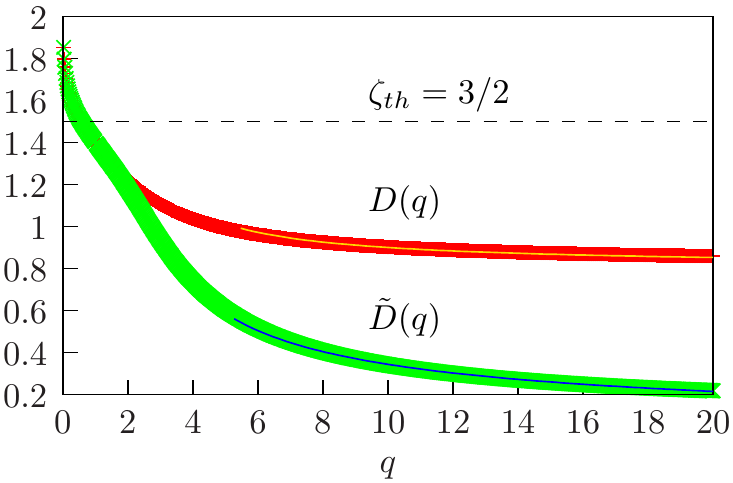}
 \caption{(color online) Generalized dimensions $D(q)$ and
$\tilde{D}(q)$ at criticality $T=T_{c,\kappa}$.
 The dashed line is  $D_{T>T_c}=3/2$.
Solid lines are monomial fits for large q,
see text.}
\label{fig:gendim}
\end{center}
\end{figure}

In Fig.\ \ref{fig:gendim}, we present numerical results 
for $D(q)$ and $\tilde{D}(q)$ for  a SDL in disorder at criticality. 
As a control for the numerics one can use the information dimension $D(1)$,
which should coincide with its high temperature value
$\left.D(1)\right|_{T>T_c}=\chi=\zeta d=3/2$ at criticality. 
The data appears to resemble
the Anderson transition-like scenario that has been reported for the
DL \cite{Monthus2007} with $D(q)=\tilde{D}(q)$ for $q<q*\approx 1.5$. The
separation of $D(q)$ and $\tilde{D}(q)$ indicates different behaviors
 of typical and average values of $Y_{q>q*}$, from
which one is tempted to conclude that these quantities are {\em not}
self-averaging. Furthermore both
$Y_q^{av}$ and $Y_q^{typ}$ are diverging faster than exponential for $q<0$,
which leads to $D(q<0)=\tilde{D}(q<0)=\infty$. 
The information dimension $D(1)$ is measured to be about $1.4$
 and, therefore, does not coincide with the expected
high temperature value $3/2$. However, this could be a
 numerical artifact from limited system sizes.

\begin{figure}[t]
\begin{center}
 \includegraphics[width=0.45\textwidth]{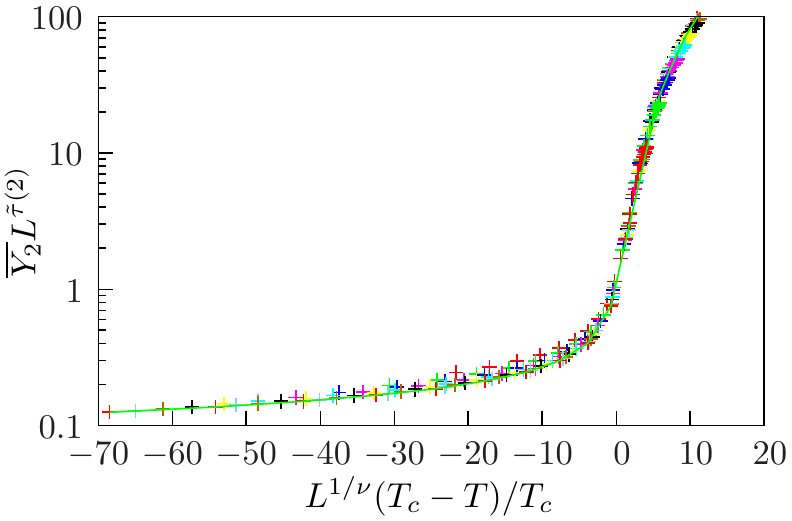}
 \caption{Finite size scaling
$\overline{Y_2}L^{\tilde{\tau}(2)}((T_c-T)/T_c \cdot L^{1/\nu})$ giving
$\nu=2$
and $T_c=1.41429$ for lengths $L=20,30,\ldots,100$, the points for $L=100$ are
interconnected by a line as a visual guidance. This does not work for the
alternatively proposed value $\nu=4$. For $T<T_c$,
$\overline{Y_2}$ remains finite for large $L$ and,
therefore, the finite size scaling observable diverges as
$L^{\tilde{\tau}(2)}$.}
\label{fig:y2finscal}
\end{center}
\end{figure}

For Anderson transitions,
the finite-size scaling of the $Y_q$ at criticality 
does involve the multifractal spectrum but only one 
correlation length exponent
$\nu$ \cite{Monthus2006a,Huckestein1995},
\begin{equation}
 \overline{Y_q} = L^{-\tilde{\tau}(q)} f((T_c-T)L^{1/\nu}) .
\end{equation}
Thus, it allows for a completely independent validation (cf.\
Fig.\ \ref{fig:y2finscal}) of the critical temperature 
and the correlation length exponent,
yielding  a more exact value 
$T_c\approx 1.44$ compatible with our result from 
sections \ref{sec:existence} and \ref{TcT2}  and unambiguously a value
$1/\nu\approx 0.5$ for 
the correlation length exponent. 

\begin{figure}[t]
\begin{center}
 \includegraphics[width=0.45\textwidth]{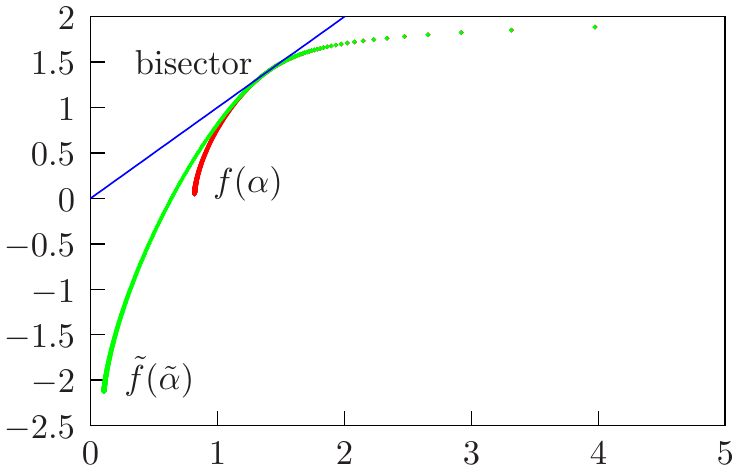}
 \caption{(color online) The singularity spectrum $f(\alpha)$, see eq.\
\eqref{eq:falpha}. For a better comparison with the expectations the plot also
shows lines corresponding to $f(\alpha)=D(2)=2$, $f(\alpha)=D(1)$ and
$f(\alpha)=\alpha$, see text. The bisector and $f(\alpha)$ touch at an
$\alpha$ that is slightly smaller than $\alpha=1.5$. We consider this to be
wrong, but it is consistent with the determined generalized dimensions,
 see also
Fig.\ \ref{fig:alphafalpha}.}
\label{fig:singspec}
\end{center}
\end{figure}

\begin{figure}[t]
\begin{center}
 \includegraphics[width=0.45\textwidth]{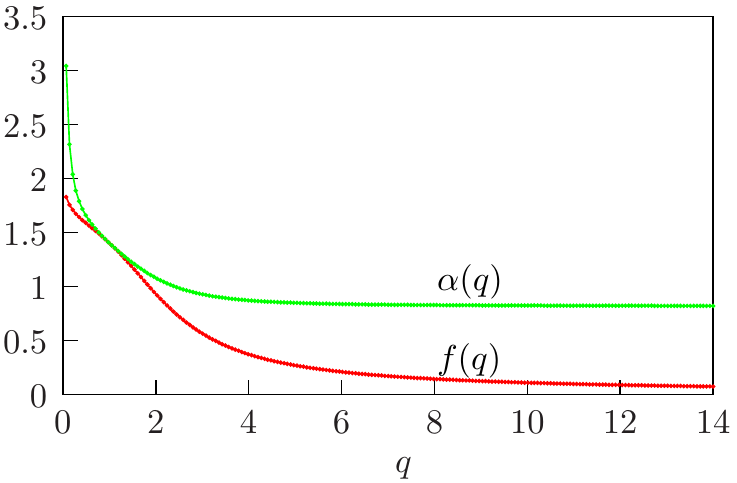}
 \caption{(color online) The directly measurable
$\alpha(q)$ (green) and $f(q)$ (red) from which Fig.\ \ref{fig:singspec}
was created. The contact point is at
$\alpha=f(\alpha)\approx 1.4$}\label{fig:alphafalpha}
\end{center}
\end{figure}

%HERE

An equivalent description of the multifractal nature is related to the
Legendre transform $f(\alpha)$ of $\tau(q)$ given by
\begin{subequations}
\label{eq:legendre}
\begin{align}
 q&=f'(\alpha) \\
 \tau(q)&=\alpha q - f(\alpha). 
\end{align}
\end{subequations}
The function $f(\alpha)$ is called the {\em singularity spectrum}, because it
gives the number $N(\alpha)\sim L^{f(\alpha)}$ of points $z$, where the weight
$w(z)$ has a singularity $w\sim L^{-\alpha}$. A Legendre transform of the
measured $\tau(q)$ and $\tilde{\tau}(q)$ would require an 
analytical continuation and, thus, be very
error-prone. Fortunately, it is possible to directly measure
$\alpha(q)$ and $f(q)$  \cite{Chhabra1989}  and, thus,
 the singularity spectrum $f(\alpha)$ parametrically:
\begin{subequations}
\label{eq:falpha}
\begin{align}
 f(q)&= - \lim_{L\rightarrow\infty} \overline{\sum_i \mu(q,z_i)
\ln{\mu(q,z_i)}}/\ln{L} \\
 \alpha(q) &= - \lim_{L\rightarrow\infty}  \overline{\sum_i \mu(q,z_i)
\ln{w(z_i)}}/\ln{L}
\end{align}
with 
\begin{align}
 \mu(q,z_i) &\equiv w^q(z_i)/(\sum_j w^q(z_j)).
\end{align}
 This method of computing
$f(\alpha)$ gives the Legendre transform of $\tau(q)$, because eq.\
\eqref{eq:legendre} implies (omitting the limits)
\begin{align}
 \tau(q)&= \alpha(q) q - f(q) \nonumber\\
	%&= - \overline{\sum_i \mu(q,z_i) ((q{-}q)\ln{w(z_i)}{+}
%\ln\sum_j w^q(z_j))}/\ln{L}\nonumber\\
	%&= - \overline{\sum_i (w^q(z_i)/\sum_k w^q(z_k))\ln\sum_j
%w^q(z_j)}/\ln{L}\nonumber\\
	&= - \overline{\ln{\sum_i w^q(z_i)}}/\ln{L}\\
        &= - \overline{\ln{Y_q(L)}}/\ln{L}\nonumber
\end{align}
\end{subequations}
in agreement with eq.\ \eqref{eq:yqtyp},  and, therefore,
 $f(\alpha)$ corresponds
to typical values of $Y_q$. Note that $\alpha(q)=\tau'(q)$ is fulfilled by
construction. This is the common definition of the (multifractal) singularity
spectrum that is also applicable for non-disordered systems. 

Here, disorder is
relevant and we need to capture not only the typical, but also the
(differing, cf.\ Fig.\ \ref{fig:gendim}) average behavior. 
Analogously to eq.\ \eqref{eq:falpha} we derive
the following computation of the Legendre transform
$\tilde{f}(\tilde{\alpha})$ of $\tilde{\tau}(q)$ (we distinguish
 between $\alpha$ and $\tilde{\alpha}$ to avoid ambiguities 
as both are directly measured). We use eq.\ \eqref{eq:gendim}
and the inverse transform of \eqref{eq:legendre}
\begin{subequations}
\begin{align}
 \tilde{\tau}(q)&=-\ln{\overline{\sum_i w^q(z_i)}}/\ln{L}\\
 \tilde{\alpha}(q)&= \tilde{\tau}'(q)\nonumber\\
 &= -  \overline{\sum_i
 w^q(z_i)\ln{w(z_i)}}/(\ln{L}\cdot\overline{\sum_i w^q(z_i)})\\
 \tilde{f}(\tilde{\alpha})&=\tilde{\alpha}q-\tilde{\tau}(q)\\
&=-\overline{\sum_i w^q(z_i)(\ln{w^q(z_i)}/\overline{\sum_j
w^q(z_j)}-1)}/\ln{L}\nonumber.
\end{align}
\end{subequations}
The spectrum $f(\alpha)$ is shown in Fig.\
\ref{fig:singspec}. Its shape matches the expectations that origin from general
properties and the known results for
the DL \cite{Monthus2007} and is consistent with the results for the Legendre
transform $\tau(q)$. Our results show that $f(\alpha)$ is a monotonic
function starting at a finite
$\alpha_\text{min}=D_{\text{min}}(q)\approx 0.8$ which is close to the DL value
$\alpha_\tau^{min}\approx
0.77$ and ending at $\alpha_{max}=\infty$, which
corresponds to the infinitely large values $\tau(q)$ for $q<0$. The maximum
value of $f(\alpha)$ is $f(\alpha\rightarrow\infty)\rightarrow D(0)$ and it
touches the bisector  $f=\alpha$ 
around $\alpha=D(1)\approx 1.4$, thus confirming the
previously found deviation $D(1)\neq 3/2$. We see no
indication that $f(\alpha)$ becomes
negative somewhere, which would describe rare events (number
decreases exponentially in $L$), but this is to be expected as $f(\alpha)$
contains the {\em typical} behavior and should not become
negative\cite{Monthus2007}. The {\em average} behavior that leads to
$\tilde{f}(\tilde{\alpha})$ does include rare events and does not seem to have
a finite $\alpha_{min}$, implying $\tilde{D}(\infty)=0$. We back this by noting
that the we can achieve a good
fit of the data for $\tau(q)$ and $\tilde\tau(q)$ at large q (we used $5<q<20$)
with an monomial Ansatz $f(q)=aq^b$. For an approximate analysis
we round $b$ to one decimal giving
\begin{align}
 \tau(q\gg 1)&\approx aq\\
 \tilde\tau(q\gg 1)&\approx\tilde{a}q^{2/5}
\end{align}
with some constants $a,\tilde{a}$. We apply the Legendre transform
and get for $q\gg1$
\begin{alignat*}{3}
 \alpha(q)&\approx a \approx 0.81& \qquad
 f(\alpha)&\approx 0\\
 \tilde\alpha(q) &\sim q^{-3/5} \rightarrow 0  & \qquad
 \tilde f(\tilde\alpha) &\sim - \tilde\alpha^{-2/3} \rightarrow -\infty
\end{alignat*}

In summary, we see a (Anderson transition-like) scenario in the
multifractal analysis of the localization transition of a SDL in 1+1
dimensions which is very similar to the findings in Ref.\ \cite{Monthus2007} for
a DL in
1+3 dimensions despite obvious differences due to the different geometry. 
In particular, the multifractal structure of the statistical weights differs
between typical and average values both for DLs and SDLs.   This becomes
apparent in
the generalized dimensions $D(q), \tilde{D}(q)$, which differ for $q\gtrapprox
1.5$. We also find a matching critical correlation length exponent $\nu$.
Additionally, we showed that the average behavior is significantly influenced by
rare events, leading to negative values in $\tilde{f}$, the Legendre transform
of $\tilde{D}(q)$. The significance of rare (extreme) events at criticality is
in accordance with our findings for the energy distribution.

%HERE

%%%%%%%%%%%%%%%%%%%%%%%%%%%
%Conclusion
\section{Conclusion}

We studied stiff directed lines (SDLs)  in 1+d dimensions subject to 
 quenched  short-range random potential analytically and numerically. 
Using Flory-type scaling 
 arguments and a replica calculation we show that,
in dimensions $d>2/3$,  a localization transition exists 
from a high temperature
phase, where the system is essentially annealed, to a
disorder dominated low temperature phase. 
 The low temperature  phase is characterized
by large  free energy fluctuations with  an exponent 
$\omega>0$ and a 
roughness exponent $\zeta$, which slightly exceeds the 
thermal value $\zeta_{th}=3/2$ for a SDL.
Flory arguments suggest $\zeta = 7/(4+d)$ for a SDL. 
Both exponents are related by $\omega= 2\zeta -3$.  

 We find a reduction of the 
persistence length of a stiff directed line by disorder. In the low
temperature phase, the persistence length is   disorder-induced 
and  temperature-independent.

We also performed variational replica  functional 
renormalization group (FRG) calculations. The replica approach 
gives no conclusive results on the exponents for $d>2/3$ but 
supports the existence of a localization transition. 
The FRG calculation is performed  for D+d dimensional 
manifolds governed by bending energies 
with the SDL corresponding to D=1 and 
employs an expansion in 
$\epsilon = 8-D$ dimensions. For a SDL in 1+d dimensions,
the FRG result suggests the existence of an upper critical dimension 
$d_u <1$, which we can rule out by numerical calculations. 

For the SDL in 1+1 dimensions we performed
 extensive numerical transfer matrix 
calculations and  find the existence of a localization transition 
and an  exponent
 $\omega\approx 0.18$  in the low temperature 
disorder dominated phase. The value for $\omega$ 
is close to the  established value $\omega \approx 0.186$ 
 for  directed lines (DLs) under
tension in 1+3 dimensions. Moreover, 
 the rescaled free energy distributions  are {\em identical}. 
Both points suggest that the nature of
the low-temperature phase is very similar, if not identical. 

The multifractal analysis reveals a very similar structure of the
statistical weights at the critical temperature for DLs in 1+3 dimensions and
SDLs in 1+1 dimensions. It also allows for a
determination of the correlation length exponent $\nu$, for which we find,
again in accordance with the findings for a DL in 1+3 dimensions, $\nu=2$.
Additionally, we find evidence for the relevance of rare events at criticality.

\begin{figure}[t]
 \begin{center}
  \includegraphics[width=0.45\textwidth]{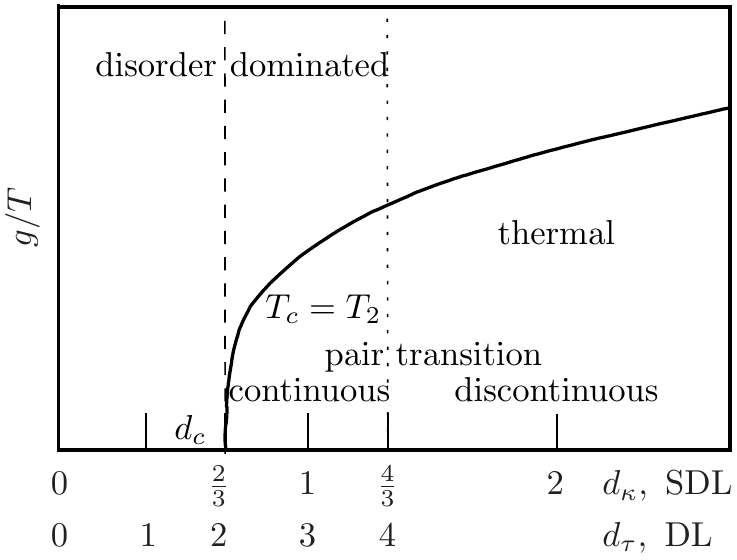}
 \caption{Schematic summary of our results supporting  a relation between 
   DLs in 1+3d and SDLs in 1+d dimensions and the importance 
  of replica pair interactions for the localization transition
  in a random potential.}\label{fig:result}
 \end{center}
\end{figure}

This strongly supports a relation between DLs in 1+3d
and SDLs in 1+d dimensions, which is based on identical  return 
exponents $\chi$ for two replicas to meet. The validity of 
a relation based on properties of a single replica pair suggests that 
the critical properties of  DLs in a short-range 
random potential are governed by replica pair interactions.
The mapping can make 
 DL transitions in high dimensions  computationally accessible,
 which we demonstrated in showing that the two-replica overlap 
provides a valid order parameter across the localization transition 
of SDLs in 1+1 dimensions. 
Furthermore, the importance of pair interactions suggests
 that the critical temperature for DLs in random potentials 
is  indeed 
identical to the temperature below which the
ratio of the second moment of the partition function and the square of its
first moment diverges, which implies that 
the localization transition temperature $T_c$ in a random potential equals 
  the replica pair transition temperature $T_2$ for  replica pair binding.
The equality $T_2=T_c$ has been originally put forward and verified 
numerically for DLs in 1+3 dimensions \cite{Monthus2006a, Monthus2007}.
 Using the numerical transfer matrix approach we could verify this 
conjecture also for
SDLs in 1+1 dimensions.
Our findings are summarized schematically  in Figs.\
\ref{fig:result2} and \ref{fig:result}.

The binding transition of 
DL pairs becomes discontinuous for $d>4$  
and, analogously, the binding of  SDL pairs
for $d>4/3$  \cite{Kierfeld2003,Kierfeld2005}. 
Because DLs in random potentials are 
equivalent to  the KPZ equation \cite{Kardar1986}, the validated relation to
the SDL suggests that the roughening 
transition of the KPZ problem could acquire similar  
discontinuous features for $d>4$ dimensions.  Thus,
 $d=4$ would remain a
special dimension, although it is not the upper critical dimension
\cite{Marinari2002}.

%%%%%%%%%%%%%%%%%%%%
\section{Acknowledgments}

We acknowledge financial support by  the Deutsche Forschungsgemeinschaft
(KI 662/2-1).

%%%%%%%%%%%%%%%%%%%%%%%%%%%%%%%%%%%%%%%%%%%%%%%%%%%%%%%%%%%%%%%%%%

\end{document}